\documentclass[reprint,
 amsmath,amssymb,
 aps,
superscriptaddress]{revtex4-2}
\usepackage{bm, color}

\usepackage[colorlinks = true,
            allcolors = blue]{hyperref}
\usepackage{placeins}
\usepackage{float}
\usepackage{hyperref}
\usepackage{braket}
\usepackage{xcolor}

\usepackage{graphicx}
\usepackage{dcolumn}
\usepackage{bm}
\usepackage{natbib}

\usepackage{color,soul}

\begin{document}

\preprint{APS/123-QED}
\title{Performance of  wave function and Green’s function methods for non-equilibrium many-body dynamics}

\author{Cian C. Reeves}
\affiliation{%
Department of Physics, University of California, Santa Barbara, Santa Barbara, CA 93117
}%

\author{Gaurav Harsha}
\affiliation{Department of Chemistry, University of Michigan, Ann Arbor, Michigan 48109, USA}

\author{Avijit Shee}
\affiliation{Department of Chemistry, University of California, Berkeley, USA}

\author{Yuanran Zhu}
\affiliation{Applied Mathematics and Computational Research Division, Lawrence Berkeley National Laboratory,
Berkeley, CA 94720, USA}
\author{Thomas Blommel}
\affiliation{%
Department of Chemistry and Biochemistry, University of California, Santa Barbara, Santa Barbara, CA 93117
}%
\author{Chao Yang}
\affiliation{Applied Mathematics and Computational Research Division, Lawrence Berkeley National Laboratory,
Berkeley, CA 94720, USA}

\author{K Birgitta Whaley }
\affiliation{Department of Chemistry, University of California, Berkeley, USA}
\affiliation{Berkeley Center for Quantum Information and Computation, Berkeley}

\author{Dominika Zgid}
\affiliation{Department of Chemistry, University of Michigan, Ann Arbor, Michigan 48109, USA}
\affiliation{Department of Physics, University of Michigan, Ann Arbor, Michigan 48109, USA}
\author{Vojt\ifmmode \check{e}\else \v{e}ch Vl\ifmmode \check{c}\else \v{c}ek}
\affiliation{%
Department of Chemistry and Biochemistry, University of California, Santa Barbara, Santa Barbara, CA 93117
}%
\affiliation{%
Department of Materials, University of California, Santa Barbara, Santa Barbara, CA 93117
}

\date{\today}
\begin{abstract}
Theoretical descriptions of the non-equilibrium dynamics of quantum many-body systems essentially employ either (i) explicit treatments, relying on the truncation of the expansion of the many-body wavefunction, (ii) compressed representations of the many-body wavefunction, or (iii) evolution of an effective (downfolded) representation through Green’s functions.
 In this work, we select representative cases of each of the methods and address how these complementary approaches capture the dynamics driven by intense field perturbations to non-equilibrium states. Under strong driving, the systems are characterized by strong entanglement of the single particle density matrix and natural populations approaching those of a strongly interacting equilibrium system. We generate a representative set of results that are numerically exact and form a basis for a critical comparison of the distinct families of methods. We demonstrate that the compressed formulation based on similarity-transformed Hamiltonians (coupled cluster approach) is practically exact in weak fields and, hence, weakly or moderately correlated systems. Coupled cluster, however, struggles for strong driving fields, under which the system exhibits strongly correlated behavior, as measured by the von Neumann entropy of the single particle density matrix. The dynamics predicted by Green's functions in the (widely popular) $GW$ approximation are less accurate, but improve significantly upon the mean-field results in the strongly driven regime.

\end{abstract}

\maketitle

\section{\label{Intro}Introduction}
The properties of non-equilibrium quantum systems have drawn great attention in recent years.  Driving a system can alter its properties, leading to phase changes\cite{Beebe_2017,Disa_2023}, exotic states of matter\cite{Dong_2021,Bao_2022} and allowing for tuning of material properties through continuous driving\cite{Nuske_2020,Zhou_2023}.  Motivated by new experimental techniques and observations, the theoretical study of these non-equilibrium systems is increasingly attracting wide attention\cite{arponen_variational_1983,ishikawa_review_2015,li_real-time_2020,Sun_2021,Karlsson_2021,Perfetto_2022,sverdrup_ofstad_time_dependent_2023,Joost_2022,Farzanehpour_2012,Aryasetiawan_2002,Karlsson_2011,CAPELLE201391}. This has led to a range of theoretical techniques that have found success for equilibrium problems, to be extended to the non-equilibrium regime.  
The available \textit{ab initio} techniques that are systematically improvable employ either a wave function or the many-body Green's function as the basis of their solution\cite{martin_2016}. The former are typically used in quantum chemistry, while the latter is a suitable framework for condensed matter physics problems. There is however lack of critical assessment of their most appropriate domain applicability across various non-equilibrium regimes. While benchmarks of these \textit{ab initio} methods do exist\cite{sverdrup_ofstad_time_dependent_2023,pathak_time-dependent_2020,pathak_2020,pathak_2021,Hermanns_2014,Schlunzen_2016,Schlunzen_2016_2,Schlunzen_2017,Reeves_2023_2,Schmitt_2010,von_friesen_2010_2,von_Friesen_2009}they tend to focus on wave function based approaches \emph{or} Green's function based approaches but not both, {an exception to this statement are the benchmarks of TD-DFT and NEGFs in lattice models presented in Refs. \cite{almbladh_2016,verdozzi_2011}. However, to our knowledge a comprehensive comparison between the methods presented here has not been performed. } While these approaches are, to a large extent, complementary formulations of the time dependent problem, they practically tackle the many-body description differently and hence face distinct challenges which are explored in this paper. This study will help understand the respective limitations of each scheme, especially relative to one another, highlighting what problems each can most effectively be applied to.  This understanding is crucial for creating a multiscale framework that leverages multiple time-dependent methods to tackle wide ranging problems.

Among wave function methods, the time dependent exact diagonalization (TD-ED) is equivalent to the time evolution of the Schr\"odinger equation and hence it provides benchmark results. Due to its cost, it is applicable to only small finite systems. In the quantum chemistry community, it is  referred to as time dependent full configuration interaction (TD-FCI)\footnote{Note, that in ED the system Hamiltonian is diagonalized providing all the eigenvectors, while in TD-FCI customarily only few eigenvectors are obtained.}. The time dependent methods based on an approximate (truncated) CI expansion enjoy much success, for example, time-dependent configuration interaction singles (TD-CIS) \cite{TDCIS_Santra_PRA2006} method is very well suited for ionization dynamics. However, such methods have very limited applications, and active space-based CI methods are often preferred. One of the methods in that direction is the extension of multi-configurational time dependent Hartree approaches (MCTDH)~\cite{RevModPhys.92.011001} to atomic problems such as ionization of helium~\cite{10.1063/1.3553176}. Several other methods in the same vein are also known, such as the time-dependent restricted active space CI (TD-RASCI) \cite{TDRASCI_BonitzPRA2012}, and the complete active space self-consistent field (TD-CASSCF) \cite{TDCASSCF_Sato2013}. The computational cost of those techniques scales exponentially with the size of the Hilbert space, each particular formulation has a different scaling prefactor, but are generally limited to small systems.

The reduced (non-exponential) scaling wavefunction-based approaches leverage some form of compressed representation, e.g., as time dependent density matrix renormalization group method (TD-DMRG)~\cite{Daley_2004,PhysRevLett.93.207205,SCHOLLWOCK201196,RevModPhys.77.259} or time dependent formulation of coupled cluster~\cite{crawford_introduction_2000,bartlett_coupled-cluster_2007} (TD-CC) methods.
For these approaches, the price of the reduction in computational scaling comes with the establishment of the domain of most suitable applicability. TD-DMRG is usually most successful in 1D systems, while TD-CC, following the ground-state version, is typically suitable for a wide variety of weakly correlated problems.~\cite{walz_application_2012, koulias_relativistic_2019, skeidsvoll_time-dependent_2020, skeidsvoll_comparing_2023} 
One of the earliest formulations of TD-CC, of particular interest to this work, was proposed by Arponen~\cite{arponen_variational_1983}.
In recent years, new developments in theory and implementation have emerged.~\cite{kvaal_ab_2012, sato_communication_2018, pathak_time-dependent_2020, peng_conservation_2021}

In contrast to the wavefunction methods, the methods applied to condensed matter systems are most conveniently based upon the Green's function (GF) formulation of the many-body perturbation theory (MBPT)\cite{ Reining_2018,Onida_2002,Aulbur_2000}. This approach recasts the many-body problem onto a set of effective correlators, most typically the one-quasiparticle GF\cite{martin_2016}, capturing the probability amplitude associated with the propagation of a quasiparticle.  As an effective single particle quantity, GF MBPT can be efficiently implemented\cite{Wilhelm_2021,Neuhauser_2014} treating systems with thousands of electrons \cite{Vlcek_2018,Gao2016,Govoni_2015}.  The complexity is accounted for via the space-time nonlocal self-energy that is systematically built by expansion in the system fluctuations\cite{Hedin_1965,vlcek2019stochastic,Mejuto_2022}, making GF MBPT one of the most powerful methods for studying electronic phenomena in large scale systems.  It is formally straightforward to extend GF MBPT to non-equilibrium settings via the set of integrodifferential Kadanoff-Baym equations (KBEs)\cite{stefanucci2013nonequilibrium,kadanoff_2018,schwinger_1961,Keldysh_1964}. The cost of solving KBEs scales cubically with the number of timesteps\cite{bonitz2015quantum}. Despite this the KBEs and approximations to the KBEs have been applied to study a range of non-equilibrium systems\cite{Kwong_1998,Kwong_2000,Lake_1992,Schmitt_Rink_1998,Banyai_1998,Dahlen_2007,Kremp_1999}, however the high cost has prohibited their widespread application.  Multiple cost reduction schemes have been proposed, in particular in recent years\cite{Kaye_2021,Kaye_2023,Yin_2021,Stahl_2022,Lipavsky_1986,Schlunzen_2020,Reeves_2023,Reeves_2024}. Yet, the most widely used of these is the Hartree-Fock generalized Kadanoff-Baym ansatz (HF-GKBA)\cite{Lipavsky_1986,Schlunzen_2017,Bonitz_2013,Balzer_2013,Hermanns_2013} which can be formulated in linear scaling fashion\cite{Schlunzen_2020,Pavlyukh_2022}, even when using advance self-energy approximations\cite{Joost_2022}.  In some cases the HF-GKBA can even improve on elements of the KBEs, specifically the tendency of the KBEs to be overdamped and reach artificial steady states\cite{von_Friesen_2010,von_Friesen_2009}.

In this paper, we critically compare the distinct formulations of the time dynamics and illustrate their performance on a series of numerical benchmarks for a generalized Hubbard chain driven out of equilibrium by an external pulse. Using TD-FCI, TD-CC, the KBEs, the HF-GKBA and time dependent Hartree-Fock(TD-HF), we study the dynamics of the density matrix and study the effects of different strengths of non-equilibrium perturbation.
In general, the TD-FCI results serve as near-exact baseline against which other methods are compared.

The paper first reviews the essential parts of the distinct formulations and approximations in the theory section. The results then summarize the findings for lattice problems perturbed by pulses of increasing strengths that drive the system towards the regime which is shown to exhibit characteristics of strong correlations. We map the capacity of the distinct formalisms to capture the dynamics for those regimes and also for systems with short- and long-range interactions, with and without non-local memory effects (in the context of GF formalisms), and confirm our findings by testing systems of various sizes. The details of the theory and implementation of the methods are in the supplementary information\cite{supp}, while the main text focuses on the analysis of the results, which are summarized in the conclusion section.

\section{Theory}
In this work, we explore representative formulations of three conceptually distinct methodologies, configuration interactions (based on truncation of the wavefunction expansion in the space of configurations represented by Slater determinants), coupled cluster approach (employing the compression of the many-body wavefunction), and many-body Green's function techniques (based on a downfolded effective representation of excitation evolution). In the following subsections we present the key underlying ideas and we leave much of the technical details of each approach to the supplemental information\cite{supp}.   

\subsection{Time-dependent configuration interaction}
The time dependent full configuration interaction (TD-FCI) approach is taken as a reference point for assessing the performance of the other methods. In TD-FCI, we express the time-dependent Schr\"odinger equation in configuration space:
\begin{equation}
    i \frac {\partial C_I}{\partial t} = \sum_J H_{IJ}(t) C_J \label{eq:TDCI}
\end{equation}
where, $I$ and $J$ stand for all electronic configurations generated for a given problem, and $C_I$ and $C_J$ are the coefficients of those configurations. To solve this equation for a general time-dependent Hamiltonian we numerically implement the time-ordering and exponential of a large matrix, $H_{IJ}$, based on the Lanczos technique \cite{ParkLight_JCP86}. With this approach, we construct a propagator in a small subspace based on a reference form of the wavefunction at time $t$, i.e., $|\Psi (t) \rangle$. For example, at $t=t_0$, we build a propagator based on the ground state wave function of the time-independent Hamiltonian. Such a propagator is however valid only for a short time. Through a regular renewed search of the ``active'' subspace we can ultimately propagate the wavefunction for long times. The procedure is numerically stable and yields unitary dynamics; the details of the implementation are provided in the supplemental information\cite{supp}. For simplicity we will still refer to this method as TD-FCI, despite the use of a small subspace to build the propagator.

\subsection{Time-dependent coupled cluster}

Coupled cluster (CC) is another formalism expanding the wavefunction as a linear combination of configurations.
In CC, one introduces a non-Hermitian approximation to expectation values $\braket{A}_{CC} = \braket{\Psi_L | A | \Psi_R}$, where the bra and ket wavefunctions are defined as
\begin{subequations}
    \begin{align}
        \ket{\Psi_{R}} &= e^{T} \ket{\Phi},
        \\
        \bra{\Psi_{L}} &= \bra{\Phi} \left( 1 + Z \right) e^{-T},
    \end{align}
\end{subequations}
where $T=\sum_\mu \tau_\mu \gamma_\mu^\dagger$ and $Z = \sum_\mu z_\mu \gamma_\mu$ are composed of particle-hole excitation $\gamma_\mu^\dagger$ and de-excitation $\gamma_\mu$ operators, defined with respect to the Slater determinant reference $\ket{\Phi}$.

In time-dependent CC, the amplitudes $\tau_\mu$ and $z_\mu$ carry all the time-dependence, and are found by making the action integral,
\begin{equation}
    \begin{split}
        S &= \int dt \mathcal{L}(t) \\&= \int dt \braket{
        \Phi | \left(1 + Z(t)\right) e^{-T(t)}
        \left(
            i \frac{\partial}{\partial t} - H(t)
        \right) e^{T(t)} |
        \Phi
    },
    \end{split}
\end{equation}
stationary with respect to variations in $\tau_\mu (t)$ and $z_\mu (t)$.
When no truncation is imposed on the operators $T$ and $Z$, CC is equivalent to exact diagonalization, i.e., equivalent to FCI (discussed above) where the wavefunction is constructed as a linear combination of all possible Slater determinants within the given spin-orbital basis.
In practice, however, truncation is necessary. Fortunately, in weak-to-moderately correlated systems, it is sufficient to include only single and double particle-hole excitation (de-excitation) operators in $T$ ($Z$).
This leads to the well known CCSD approximation which is used for all the results and discussion presented in this paper.

It is noteworthy that due to the exponential parameterization, the ket wavefunction in truncated CC includes contributions from all Slater determinants, although these contributions are factorized in terms of the lower-rank parameters.
This is in contrast with truncated CI, where the cut-off is imposed explicitly on the excitation rank of determinants.
At the same time, the asymmetric nature of CC expectation values leads to a violation of the variational principle.
Consequently, in TD-CC, real observable quantities may develop un-physical or complex parts.
However, the unphysical terms often serve as a diagnostic tool for the CCSD approximation. In the moderately interacting regimes, where CC works well, the expectation values are well behaved. The increasing magnitude of unphysical expectation values generally coincides with the onset of strong correlations, where CC is known to fail\cite{thomas_complex_2021}. This point will be discussed in the sections below.
Further details regarding the derivation and implementation of the TD-CC equations are provided in the supplemental information.~\cite{supp}

\subsection{Green's function methods}

Finally we study two approaches based on the time evolution of the one-particle Green's function (GF), i.e., the time dependent probability amplitude associated with propagation of a quasiparticle. At non-equilibrium, the GF is an explicit function of two time arguments, and their respective ordering on the real time axis.  (as well as imaginary time axis for the finite temperature formalisms). The Kadanoff-Baym equations, representing a set of theoretically exact integro-differential equations for the two-time GF, and are given by\cite{Stan_2009}
\begin{equation}\label{eq:KBE}
           \begin{split}
                [-\partial_\tau - h] G^\mathrm{M}(\tau) &= \delta(\tau) + \int_0^\beta d\Bar{\tau}\Sigma^\mathrm{M}(\tau-\Bar{\tau})G^\mathrm{M}(\bar{\tau}),\\
                i\partial_{t_1} G^{\rceil}(t_1,-i\tau) &= h^{\textrm{HF}}(t_1)G^{\rceil}(t_1,-i\tau) + I^{\rceil}(t_1,-i\tau),\\
                -i\partial_{t_2} G^{\lceil}(-i\tau,t_2) &= G^{\lceil}(-i\tau,t_2)h^{\textrm{HF}}(t_2) + I^{\lceil}(-i\tau,t_1),\\
                i\partial_{t_1} G^{\lessgtr}(t_1,t_2) &= h^{\textrm{HF}}(t)G^{\lessgtr}(t_1,t_2) + I_1^{\lessgtr}(t_1,t_2),\\
                -i\partial_{t_2} G^{\lessgtr}(t_1,t_2) &= G^{\lessgtr}(t_1,t_2)h^{\textrm{HF}}(t_2) + I_2^{\lessgtr}(t_1,t_2),\\
            \end{split}
\end{equation}

Here $\Sigma$ is a space-time nonlocal effective potential embodying all many-body interactions;  $G^{\mathrm{M},<,>,\rceil,\lceil}$ are various components of the Green's function that depend on which of the time arguments are in real time versus in imaginary time and the $I^{<,>,\rceil,\lceil}$ are integrals that account for many-body correlation effects in the system.  A full description and discussion of these quantities is given in the supplemental information\cite{supp}.  

The KBEs are approximate only through the choice of the self-energy. In this work, we focus on the $GW$ approximation, representing the arguably most popular choice for simulations of condensed matter systems. In essence, this approach accounts for dynamical density-density interactions, which dominate the correlation effects in weakly and moderately correlated systems such as semiconductors\cite{Hedin_1965,hybertsen1986electron,martin_2016}.  The combination of integrals in the equations above and their two-time nature means solving the KBEs scales cubically in the number of time steps, making these calculations prohibitively expensive.  Furthermore, the KBEs often suffer from issues of artificial damping coming from the approximate self-energy\cite{von_Friesen_2009,von_Friesen_2010}.

Given the cost of KBEs propagation, they often solved only approximately, and the most common route is to employ the Hartree-Fock Generalized Kadanoff-Baym ansatz (HF-GKBA).  It assumes a particular self-energy for the time diagonal component of the GF and takes the off-diagonal self-energy to be approximated by only the bare exchange interactions (hence the Hartree-Fock (HF) approximation in its name). By using the HF form of the time off-diagonals, it is possible to derive an ordinary differential equation (ODE) scheme for the evolution of the equal-time component of the GF. This translates to a time-linear formalism that recasts the equation of motion using the explicit form of the two particle GF instead of making use of $\Sigma$ \cite{Joost_2020,Schlunzen_2020}.  The equations of motion for this scheme are given by:
\begin{equation}\label{eq:G1-G2}
\begin{split}
       i \partial_t G^<_{ij}(t) &= [h^{\textrm{HF}}(t), G^<(t)]_{ij} + [I+I^\dagger]_{ij}(t),\\  
               I_{ij}(t)&=-i\sum_{klp} w_{iklp}(t)\mathcal{G}_{lpjk}(t),\\
       i\partial_t \mathcal{G}_{ijkl}(t) &= [h^{(2),\textrm{HF}}(t),\mathcal{G}(t)]_{ijkl} \\&\hspace{20mm}+\Psi_{ijkl}(t) + \Pi_{ijkl} -\Pi_{lkji}^*.
\end{split} 
\end{equation}
Here $\mathcal{G}(t)$ is a two particle propagator, $\Psi_{ijkl}$ accounts for pair correlations built up due to two-particle scattering events and $\Pi_{ijkl}$ accounts for polarization effects\cite{Joost_2020}. While approximate, the HF-GKBA does not suffer from artificial damping as the KBEs do and thus in some cases offers improved results over the KBEs at a much reduced cost. This further enables the use of more advanced forms of $\Sigma$, the study of larger systems, and the ability to perform longer time evolution. A more detailed description of the HF-GKBA and the ODE scheme are given in the supplemental information\cite{supp}. In this manuscript we primarily use the HF-GKBA results as our GF method. This allows us to use more advanced self-energies and do a broader study of systems and parameters.  In Sec.~\ref{sec:KBE_effects} we provide analysis of the performance of the KBEs for a subset of parameters.{It is important to note here that the HF-GKBA is a zero temperature formalism that does not invoke time-evolution along the imaginary axis.  Rather, it includes initial correlations through an adiabatic connection to a non-interacting ground state.  This can also be performed with the KBEs, eliminating the need to consider $G^{\lceil,\rceil,\mathrm{M}}$.  Both the finite and zero temperature formalism of the NEGF equations of motion can be derived from the general Keldysh NEGF formalism that serves as a base for NEGF methods. }

\subsection{Model Systems}\label{sec:model_systems}
To benchmark different time-dependent methods, we use a generalization of the one-dimensional Hubbard model, which is driven out of equilibrium with the help of an electric-field pulse. The Hamiltonian for the system is written as:

\begin{equation}\label{eq:MB_ham}
   \begin{split}
        \mathcal{H} &= -J\sum_{\langle i,j\rangle\sigma} c^{\dagger}_{i\sigma}c_{j\sigma}+ U\sum_{i}n_{i\uparrow}n_{i\downarrow} +U\gamma\sum_{i<j}\frac{n_in_j}{|i-j|}\\
        &\hspace{25mm}+V\sum_{i}(-1)^in_{i}+\sum_{ij}h^{\mathrm{N.E}}_{ij}c_{i}^\dagger c_{j}.
   \end{split}
\end{equation}
The first two terms are the usual nearest neighbor hopping and onsite interaction of the Hubbard model.  The third term turns the system into a lattice composed of two different atoms on alternating sites and serves to open up a trivial gap in the system. This is necessary since the HF-GKBA prepares the initial state using adiabatic switching.  The adiabatic theorem requires a gapped system to correctly prepare the HF-GKBA initial state. Such gapped models have been used in previous studies of the HF-GKBA\cite{Reeves_2023,Reeves_2023_2}. We have found $V=2$ is sufficiently large to make the adiabatic switching stable for the HF-GKBA state preperation.
We choose $J=1$ and express all interaction strengths in units of $J$.
Further, all the results and discussion assume half filling, i.e., one electron per site.

For each method the system is prepared in the corresponding ground state before being excited with a time dependent field modeled under the dipole approximation and given by
\begin{equation}\label{eq:h_NE}
    h^{\mathrm{N.E}}_{ij}(t) = \delta_{ij}\left(\frac{N-1}{2} - i\right)E\mathrm{e}^{-\frac{(t-t_0)^2}{2\sigma^2}}.
\end{equation}
Here $N$ is the number of lattice sites in the model, $E$ is the strength of the field, $\sigma$ determines the temporal width of the pulse and $t_0$ is the pulse midpoint.  For the following results we use $\sigma = 0.5J^{-1}$ and $t_0 = 5J^{-1}$.

\section{Results}
\subsection{Effect of excitation strength}\label{sec:diff_E}
First we investigate the effect of the strength of the perturbation by varying $E$ in equation \eqref{eq:h_NE} of our model.  This will capture the behavior of each method as we go from linear response type excitations up to strongly perturbed regimes.  We will use the electronic dipole, defined in equation \eqref{eq:dipole}, as our point of comparison between each method.  This is a useful measure as it is both experimentally relevant and it condenses information from the density matrix into a single value.
\begin{equation}\label{eq:dipole}
    p(t) = \frac{1}{N}\sum_{j=1}^{N/2}\left(\frac{N-1}{2} - j\right)[\rho_{jj}(t) - \rho_{N-j+1N-j+1}(t)].
\end{equation}
Here $\rho_{jj}(t)$ is the density on site $j$ at time $t$ and $N$ is the number of sites. We will compare results from each method for the system defined in equations \eqref{eq:MB_ham} and \eqref{eq:h_NE} with $N=12$, $U = 1.0J$ and $E=1.0J$, $2.0J$, $4.0J$ and $5.0J$.  Results for $U=1.0J$ and $E=3.0J$ are analyzed in Sec.~\ref{sec:diff_U}, as well as results for other interaction strengths.  This section focuses on the model with onsite interactions only, corresponding to $\gamma=0.0$ in equation \eqref{eq:MB_ham}. Results for $\gamma\neq 0$ are analyzed in Sec.~\ref{sec:long_range} .  For  $U=1.0$ each method captures ground states very well. Thus, this setup allows us to see how each method performs in non-equilibrium without having to worry about the effects of a poorly captured ground state. 

\begin{figure}
    \centering
\includegraphics[width=0.5\textwidth]{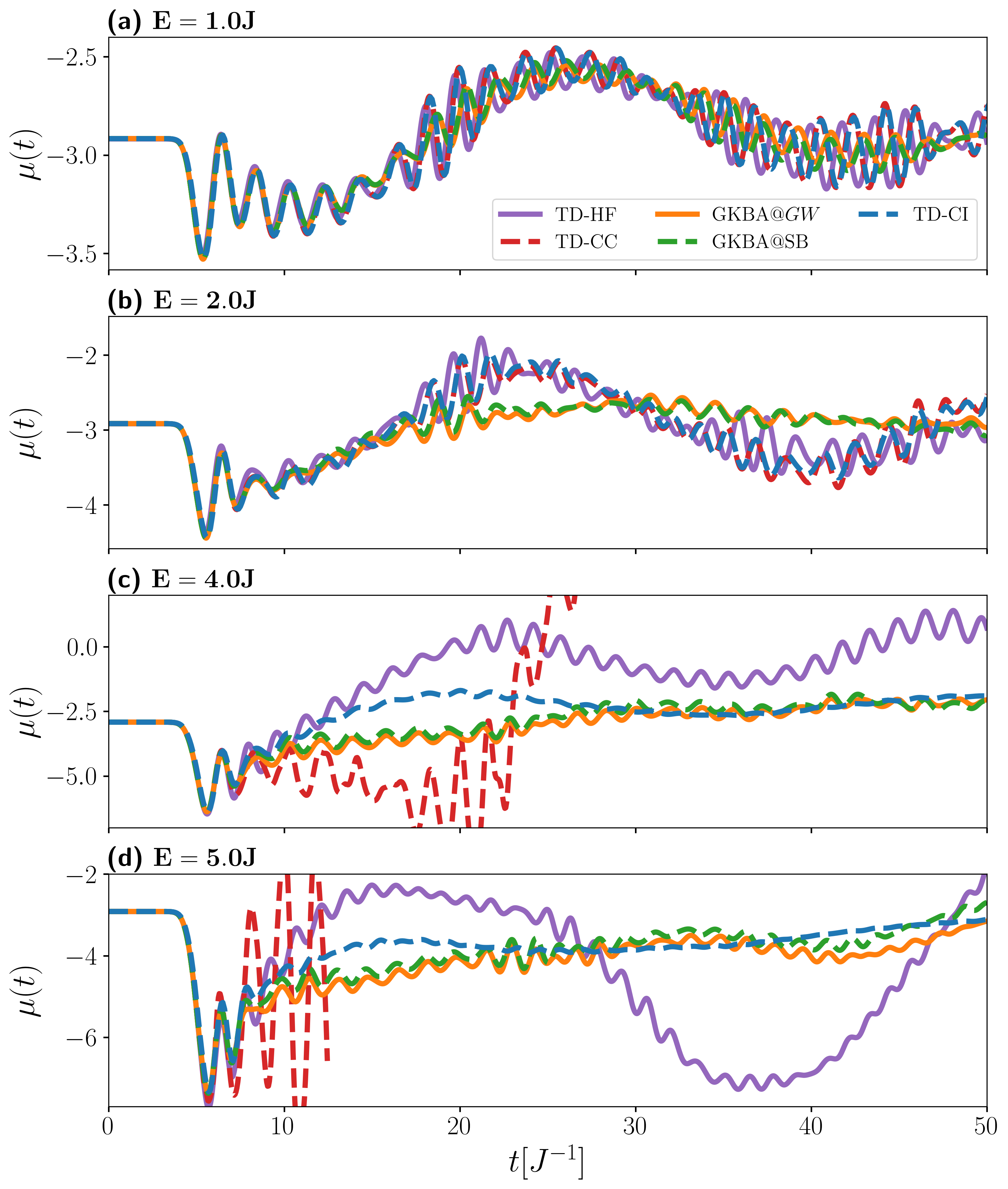}
    \caption{Dynamics of the electronic dipole moment for the generalized Hubbard model (\eqref{eq:MB_ham} and \eqref{eq:h_NE}) with on-site Coulomb interaction, for various values of electric field: a) $E=1J$, b) $E=2J$, c)$E=3J$, and d) $E=5J$.}
 
    \label{fig:diff_E}
\end{figure}

We will start this analysis by discussing the results produced by the reference TD-FCI approach, which is numerically exact and serves as our benchmark to compare our other methods against.  For $E=1.0J$, we see two dominant types of oscillation: a strong low-frequency contribution that manifests as long wavelength oscillations in the dipole and a weak high-frequency contribution.  This behavior continues with increasing $E$, however, for strong perturbation strengths, we see that the magnitude of both types of oscillations diminishes, and a yet lower frequency mode begins to dominate.  This continues until the dynamics become almost flat for $E=5.0J$ in Fig.~\ref{fig:diff_E} d).
Further results and analysis to understand these exact trends in dipole moment are discussed in Sec.~\ref{sec:exciation_analysis} and \ref{sec:discussion}, we continue this section by comparing the results produced by our other methods.

For completeness, we include the TD-HF trajectory, which serves as a baseline for the types of expansion techniques discussed below (either CC or those based on the GFs, with HF being their underlying reference Hamiltonian around which the self-energy is expanded). TD-HF is based solely on a time evolution using a single determinant. For $E=1.0J$ we see TD-HF performs well, capturing the amplitude and frequency of the oscillations of TD-FCI.  It suffers from a similar issue the HF-GKBA has in that it has a slight offset in frequency that causes it to move out of phase with the exact result as the time evolution progresses.  Interestingly for $E=2.0J$ TD-HF actually improves upon the results given by HF-GKBA, {however the frequency offset relative to TD-FCI increases compared to the $E=1.0J$ case.}  {This fortuitous improvement is likely due to the HF-GKBA improperly redistributing spectral weights into satellite solutions and will be discussed more in Sec. \ref{sec:discussion}}.  {As $E$ is increased again we see the result worsen.}  For $E=4.0J$ and $E=5.0J$ TD-HF is in agreement up to $t\approx 8J^{-1}$ but then begins to differs drastically.  The TD-HF result does not flatten out like the TD-FCI result but rather oscillates with a similar amplitude for all values of $E$ shown.  In the next paragraphs, we discuss the CC and GF-based techniques, which can be thought of as expansions around these ``correlation-free'' TD-HF dynamics.

{
The TD-CC approach is in excellent agreement with TD-FCI for the lowest two $E$ values ($E=1.0J, 2.0J$) for the entire time-evolution despite the relatively complex dynamics.  For $E=4.0J$ and $5.0J$, TD-CC captures the dynamics very closely only up until $t\approx8J^{-1}$ however soon after (around $t \approx 10 J^{-1}$), the CC approach starts to break down, the magnitude of oscillation increases significantly, and the results become unphysical. Although this is not immediately obvious from simply comparing the TD-CC and TD-FCI results for the dipole, it is shown in Fig.~S6 of the SI\cite{supp} that the TD-CC occupation numbers develop a large imaginary contribution. In fact, the disagreement of TD-CC with TD-FCI in the presence of strong perturbation can be anticipated by looking at the imaginary component of the occupation numbers.  From Fig.~S7 it is evident that even for $E=3.0J$ at long time scales, and especially for $E=4.0J, \, 5.0J$, TD-CC leads to large unphysicalities, signaling its breakdown.  At the same time, keeping track of imaginary parts or inconsistent behavior in expectation values in TD-CC can serve as a diagnostic tool to measure the efficacy of the coupled cluster approximation itself even in cases where an exact benchmark is not available.
}

We will now discuss the GF approach computed using the HF-GKBA. {A comparison of the HF-GKBA and the full KBEs will be presented in Sec.~\ref{sec:KBE_effects}}.  

{Here we consider two self-energy approximations, second Born (SB) and $GW$, due to their importance and broad application in both equilibrium  and non-equilibrium GF theory.}   {For $E=1.0J$ both self-energy approximations give excellent agreement up to $t\approx15J^{-1}$}.  After this, the HF-GKBA still shows a strong qualitative and quantitative agreement compared to the benchmark results.  {The difference between the SB and $GW$ results remains insignificant for the whole time evolution, and even as we go to larger $E$ this persists. As with TD-HF the HF-GKBA captures the low frequency mode of oscillation very well, while for longer times, the high frequency oscillations are offset compared to TD-CC and TD-FCI. Interestingly the direction of the phase shift for HF-GKBA is the opposite to that of TD-HF.}  In addition to this offset the magnitude of the high frequency oscillations are also diminished relative to the benchmark results.   {For $E=2.0J$ the HF-GKBA (for $GW$ and SB self-energies) begins to flatten out compared to TD-FCI after around $t=20J^{-1}$ but appears to be centered around the main low frequency oscillations of TD-FCI. The HF-GKBA follows the same trend as TD-FCI, however it causes the dipole oscillations to flatten faster leading to a relatively poor match around $E=2.0J$. As $E$ is increased to $4.0J$ the HF-GKBA result remains similarly flattened and now TD-FCI has also flattened significantly leading to HF-GKBA capturing the behavior well and improving significantly over TD-HF and TD-CC.} Turning finally to Fig.~\ref{fig:diff_E} d) with $E=5.0J$ we see the HF-GKBA again captures the qualitative behavior of the TD-FCI dipole well.  The HF-GKBA displays a slightly larger magnitude of oscillation than those of TD-FCI, {however the mean values are very close and again it offers a huge improvement over the TD-CC and TD-HF results.  Interestingly the addition of even the most simple time-non-local self-energy approximation (SB) leads to very different behavior compared to the completely time-local HF self-energy. In particular for high field strengths including self-energy effects leads to qualitatively correct predictions for strongly perturbed systems, showing the importance of these effects.}  
{\subsection{Effect of interaction strength}\label{sec:diff_U}}
{In this section we will analyze the results shown in Fig.~\ref{fig:diff_U} a), c) and e) showing the effects of increasing $U$ in equation \eqref{eq:MB_ham} with $\gamma = 0$.  Fig.~\ref{fig:diff_U} b), d) and f) correspond to the long-range interacting model with $\gamma = 0.5$ and will be discussed in Sec.~\ref{sec:long_range}. We continue to analyze the electronic dipole from equation \eqref{eq:dipole}. The results are shown for the intermediate value of $E=3.0J$ however a full set of results for different $E$ and $U$ values is shown in Fig. S5 of the supplemental material\cite{supp}.  Continuing the pattern from Sec.~\ref{sec:diff_E} we will begin by analyzing the TD-FCI result.}

{For $U=0.5J$ the TD-FCI result again shows contributions from a high and low-frequency component.  As $U$ is increased the dipole follows a similar trend to when we increased the field strength.  The magnitude of both types of oscillations diminishes and for $U=2.0J$ after around $t=25J^{-1}$ the dynamics have flattened similarly to the case with $U=1.0$ and $E=4.0J$ in Fig.~\ref{fig:diff_E} c).  This similarity in behavior for increasing $E$ and increasing $U$ will be analyzed and discussed in sections \ref{sec:exciation_analysis} and \ref{sec:discussion}.}

{TD-HF is in excellent agreement with TD-FCI for $U=0.5J$ up to $t\approx10J^{-1}$. While the agreement stays very good for the remainder of the time-evolution TD-HF slightly over-exaggerates the oscillation amplitudes at parts of the time-evolution.  As in the previous section TD-HF shows a slight offset in the oscillation frequencies and thus develops a phase shift relative the the benchmark.  As $U$ is increased we again see the most obvious failure of TD-HF is it's inability to diminish the oscillation amplitudes.  For $U=1.0J$ this is noticeable but the dynamics are still relatively close up to $t\approx30J^{-1}$.  However, for $U=2.0J$ after around $t=10J^{-1}$ the dynamics are drastically different to those of TD-FCI. }

{For $U=0.5J$ TD-CC performs the best of our approximate methods and is in perfect agreement with TD-FCI up to around $t=25J^{-1}$.  After this the two methods are still in excellent agreement however some deviation begins to develop.  Increasing the interaction strength to $U=1.0J$ increases the disagreement with TD-FCI as we expect.  The agreement is near perfect up until $t\approx10J^{-1}$ and remains good up until around $t=25J^{-1}$, where a underestimation of the low frequency oscillation amplitude is the main deviation from TD-FCI.  For $U=1.0J$ and $E=3.0J$ the occupations develop a small imaginary component in the occupations, shown in Fig.~S7 of the supplemental material \cite{supp}, which can, at least partially, account for the deviation from the benchmark result.  For $U=2.0J$ the TD-CC result only matches TD-FCI up to $t\approx 7J^{-1}$ after which the dynamics have no resemblance to the benchmark.}

{Finally we look at the HF-GKBA results for the SB and $GW$ self-energies.  For $U=0.5J$ both self-energy approximations lie on top of one another and both closely approximate the benchmark.  The HF-GKBA slightly underestimates the amplitudes of oscillation, however the frequencies are well matched over the time-evolution.  As $U$ is increased to $U=1.0J$ this feature is accentuated and the low frequency oscillations of the HF-GKBA results (for $GW$ and SB) flatten out significantly compared to TD-FCI.  For $U=2.0J$ the $GW$ and SB results begin to differ quite significantly from one another.  The SB result is close to $GW$ up to $t\approx20J^{-1}$ but both results are quite far from the TD-FCI result.  After $t\approx15J^{-1}$ the SB dipole begins to increase until around $t=35J^{-1}$ where it attains a value close to the benchmark.  The $GW$ result on the other hand remains relatively flat after $t\approx10J^{-1}$.  The qualitative behavior of the $GW$ result is correct, although with an offset value for the dipole and with larger high frequency oscillation amplitudes.  The HF-GKBA results again highlight the importance of non-local self energy effects for capturing qualitative behavior of these systems, as interaction strength is increased.}

\subsection{Long range interactions}\label{sec:long_range}
\begin{figure*}
    \centering
    \includegraphics[width=\linewidth]{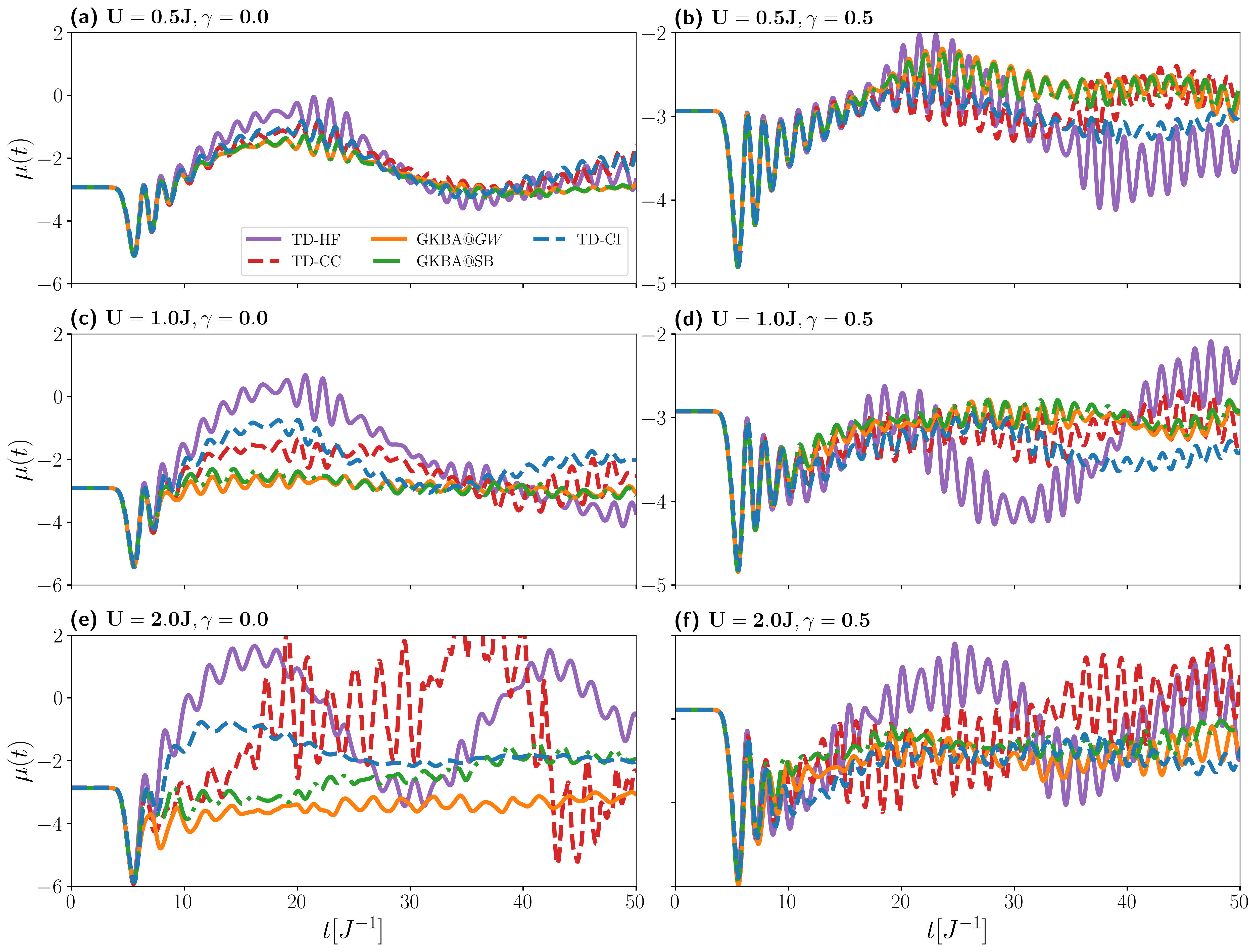}
    \caption{Dynamics of the electronic dipole moment for the generalized Hubbard model (\eqref{eq:MB_ham} and \eqref{eq:h_NE}) with electric field strength $E=3.0J$, for different values of the interaction strength with onsite interactions: a) $U=0.5J$, c) $U=1.0J$, e)$U=2.0J$, and long range interactions with $\gamma=0.5$: b) $U=0.5J$, d) $U=1.0J$, f)$U=2.0J$.}
    \label{fig:diff_U}
\end{figure*}

{The final model we will discuss has long range interactions with $\gamma = 0.5$ in equation \eqref{eq:MB_ham}. We will look at this model for $E=3.0J$ and $U=0.5J, 1.0J$ and $2.0J$.  The results are shown in Fig.~\ref{fig:diff_U} b), d) and f) and are plotted beside the onsite results for more direct comparison.  We note that the $y$-axis scale is different on the left and right of Fig.~\ref{fig:diff_U}.}

{The TD-FCI result follows a similar trend to previously.  The oscillations show a high and low frequency component and the low frequency dynamics flatten out as $U$ is increased.  Two noticeable difference between the dynamics after including long range interactions is the overall magnitude of oscillations decreases and the amplitude of the high frequency oscillations remains relatively unchanged compared to the onsite model where this amplitude decreased with increasing $U$.}

{TD-HF performs near perfectly for $U=0.5J$ up until $t\approx15J^{-1}$.  After this it exhibits the typical error of TD-HF that we have seen of over-exaggerating the oscillation amplitudes relative to TD-FCI.  Interestingly for this model TD-HF no longer shows the phase shift we have seen in the previous results, now despite the amplitudes being too large the crests and peaks of TD-HF and TD-FCI match almost perfectly for the entire time-evolution.  The agreement worsens as $U$ is increased however the dynamics of the long range model tend to be captured better by TD-HF than for the onsite model.  For $U=1.0J$ strong agreement holds until around $t=20J^{-1}$ compared to only $t\approx10J^{-1}$ in the onsite model.  For $U=2.0J$ the results agree until around $t=10J^{-1}$ before TD-HF oscillates with much larger amplitudes in both the high and low frequency components.  For $U=2.0J$ both results are too far from the benchmark for it to be obvious in which model TD-HF performs better.}

{Interestingly, for the long range model and $U=0.5$ TD-CC already begins to differ from the benchmark around $t=30J^{-1}$, sooner than for the $\gamma = 0$ model and by around $t=35J^{-1}$ it is significantly different from TD-FCI.  The level of agreement is about the same for $U=1.0J$ as for $U=0.5$.  Now the TD-CC result actually matches the TD-FCI result better than for the onsite interaction only model.  However, still after around $t=35J^{-1}$ it does not match the TD-FCI dipole any more.  For $U=2.0$ the agreement holds only up until around $t=10J^{-1}$ and after this TD-CC matches the oscillation frequencies well however the amplitudes and more precise details of the dipole dynamics are poorly captured.  The amplitudes for $U=0.5J$ and $U=1.0J$ were also over estimated, however it is much more noticeable for $U=2.0J$ perhaps pointing to a growing instability in the time evolution. However, the dynamics appear more stable and well behaved compared to those for the onsite model.}

{Turning again to the HF-GKBA, as with TD-CC we again see that for $U=0.5J$ the dynamics actually worsen compared to the onsite model.   The agreement holds only until around $t=20J^{-1}$ after which the high frequencies are still well captured but the low frequency mode flattens and is larger in magnitude than the TD-FCI result.  Again, as with TD-CC, we see that for $U=1.0J$ the HF-GBKA for both $GW$ and SB captures the initial dynamics quite closely all the way until $t=30J^{-1}$, whereas the agreement for the onsite case was only until around $t=7J^{-1}$.  For $U=2.0J$, the TD-FCI result has become more flattened out and the agreement between HF-GKBA and TD-FCI is quite good over the entire evolution.  Interestingly the SB self-energy appears to perform slightly better than the $GW$ approximation. In particular near the end of the propagation where the SB result has reduced high frequency oscillation amplitudes compared to $GW$.  Both HF-GKBA methods capture the qualitative and even quantitative behavior better than for $\gamma=0$ and their difference is also relatively minor compared to in the onsite interacting model.}

\subsection{Analysis of the KBEs}\label{sec:KBE_effects}
{Throughout this paper we have used the HF-GKBA as our method describing time-evolution with MBPT.  The HF-GKBA is known to provide results that are no worse than the KBEs with a much lower cost\cite{Reeves_2023_2}.  This lower cost not only allows us to study larger systems but also look at different self-energy approximations such as $GW$ that are difficult to study using the KBEs.  These factors have led us to using the HF-GKBA for the majority of this manuscript, however for completeness we will now look at a subset of results and see how the additional memory effects included through the KBEs modify the results presented in the previous sections.}

\begin{figure*}
    \centering
    \includegraphics[width=\linewidth]{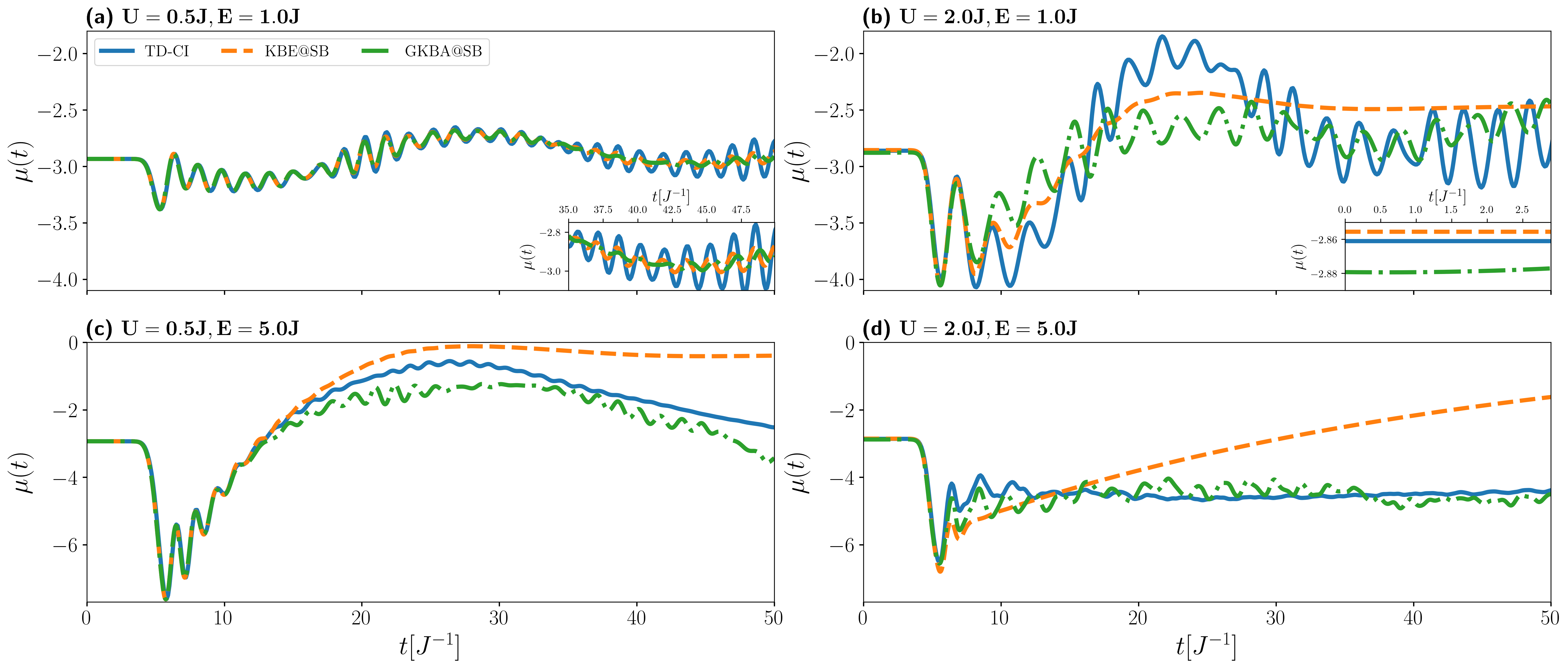}
    \caption{Comparison between HF-GKBA, TD-FCI and KBE for dynamics of the electronic dipole moment for the generalized Hubbard model (\eqref{eq:MB_ham} and \eqref{eq:h_NE}) with onsite interactions, for different values of electric field and interaction strength: a)$U=0.5J,E=1.0J$, b) $U=2.0J, E=1.0J$, c)$U=0.5J,5.0J$ and d) $U=2.0J,E=5.0J$}
    \label{fig:KBE_comparision}
\end{figure*}

{In Fig.~\ref{fig:KBE_comparision} we show results for the model with onsite interactions only.  For the long range interacting model, except for $U=0.5J$, the KBE propagation becomes unstable during the initial state preparation.  We have selected four cases to show the KBEs performance at the extrema of our parameter space. Specifically Fig.~\ref{fig:KBE_comparision} a)-d) shows the dipole from TD-FCI, HF-GKBA and the KBEs for $(U=0.5J,\ E=1.0J),\ (U=0.5J,\ E=1.0J),\ (U=2.0J,\ E=1.0J),\ (U=0.5J,\ E=5.0J)$ and $(U=2.0J,\ E=5.0J)$.  Due to the high cost of the KBEs and the $GW$ approximation, as well as the general similarity between SB and $GW$ results for the models studied here, we only show results for the SB self-energy.}

{For $U=0.5J$ and $E=1.0J$ both HF-GKBA and the KBEs capture the benchmark result perfectly up to $t\approx 30J^{-1}$.  After this both methods exhibit a decrease in the high frequency oscillations however they still closely follow the exact result.  The inset shows a zoomed in portion from $t=35J^{-1}$ to $t=50J^{-1}$.  In this inset we can see the KBEs improve on the HF-GKBA in two ways. First, the decrease in amplitude is more pronounced in the HF-GKBA and second the HF-GKBA has a slight offset in the high frequency oscillation leading to a phase shift in the dipole which is corrected by the KBEs.}

{Moving to the other panels we see the KBEs suffer from the commonly observed issue of being heavily overdamped.  For $U=2.0J$ and $E=1.0J$ the KBE results matches the benchmark until $t\approx 8J^{-1}$ but by $t\approx25J^{-1}$ the dipole has become almost completely stationary while the benchmark result still has significant oscillations.  The steady state reached by the KBEs is reasonably close to the mean value of the TD-FCI result but the HF-GKBA result, which continues to oscillate, is closer to the benchmark.  The inset shows that the KBEs slightly improve the equilibrium starting point compared to the HF-GKBA. }

{Moving in the opposite direction to $U=0.5J$ and $E=5.0J$ we see a similar outcome where the KBEs follow the benchmark up to around $t=15J^{-1}$ but then become almost completely damped. Furthermore, now the steady value the KBEs reach is far from the TD-FCI result and we see an even clearer improvement by HF-GKBA over the KBEs.}  

{For the most extreme case, $U=2.0J$ and $E=5.0J$, the KBEs now match only up until $t\approx 8J^{-1}$. After this the KBE dipole increases monotonically with no visible oscillating behavior.  This is very different from the TD-FCI dipole which reaches a near steady state modulated with small oscillations in the dipole.  Interestingly, we see that in this case although the HF-GKBA oscillates with a slightly larger amplitude, the result gives a good approximation to the TD-FCI result and again offers a clear improvement over the KBE result.}

\subsection{Exciting to strongly correlated non-equilibrium regime}\label{sec:exciation_analysis}
In this section, we further investigate the results discussed in Sec. \ref{sec:diff_E}.  As the TD-FCI results provide access to the (nearly exact) wavefunction trajectory, we use it to analyze the time evolution in greater depth. In the next section (Sec.~\ref{sec:discussion}), we rely on this analysis to illustrate the origins of failure in each method in capturing the proper physics of non-equilibrium systems. {Furthermore, we use this to help understand the observation that increasing $U$ or $E$ results in a similar trend in the dipole dynamics.}

\begin{figure*}
    \centering\includegraphics[width=\textwidth]{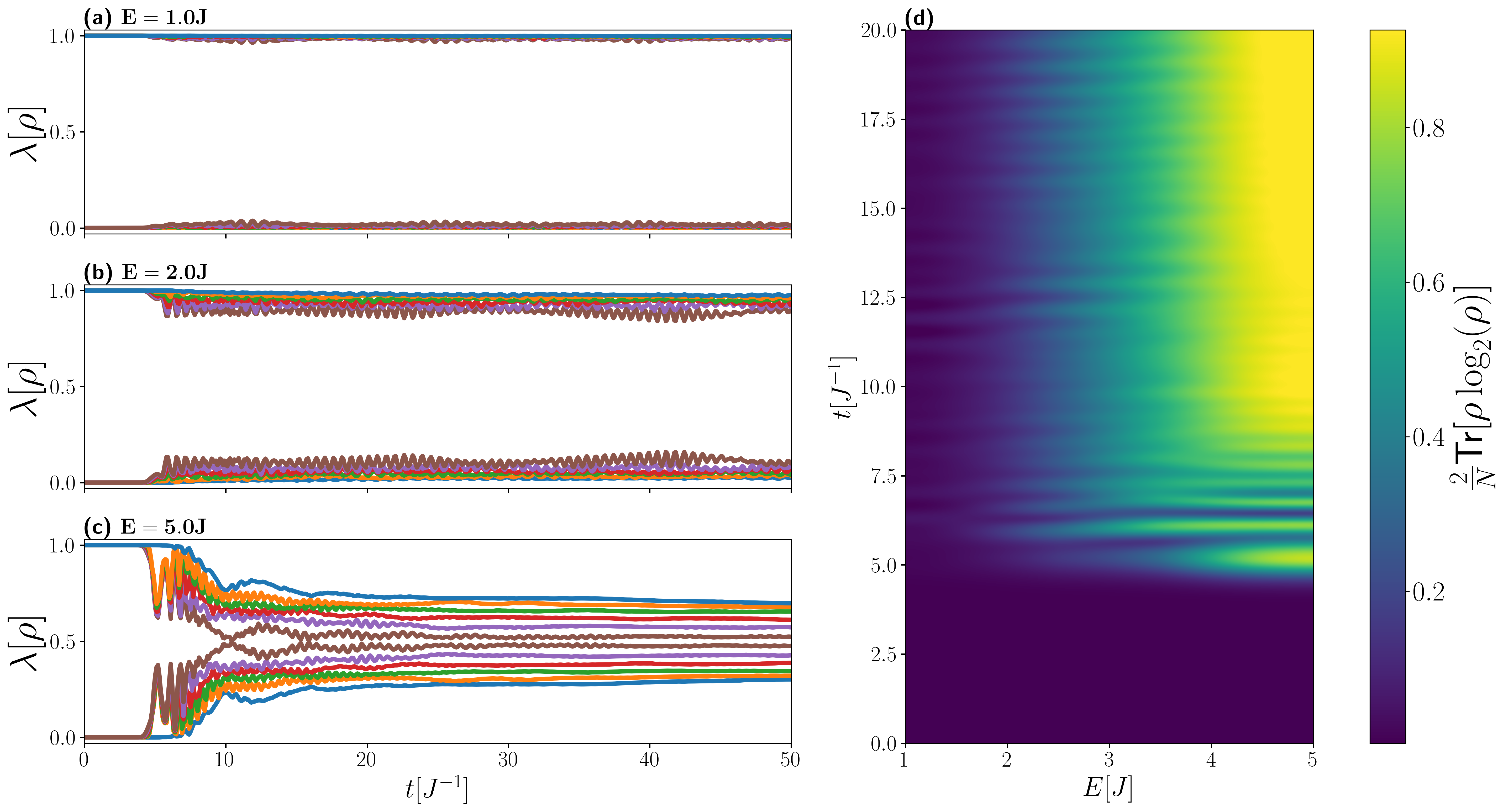}
    \caption{{\em Left}: Natural occupations of the single particle density matrix taken from TD-FCI for the model discussed in Sec.~\ref{sec:model_systems} and Sec.~\ref{sec:diff_E} for a) $E=1.0J$, b) $E=2.0J$ and c) $E=5.0J$. {\em Right:} The entropy for a scan of $E$ values from $1.0J$ to $5.0J$}
    \label{fig:entropy}
\end{figure*}

In Fig.~\ref{fig:entropy}(a)-(c), we show the natural occupations of the single particle density matrix  for the model discussed in Sec.~\ref{sec:model_systems} and Sec.~\ref{sec:diff_E} for $E=1.0J$, $E=2.0J$ and $E=5.0J$.  The natural occupations correspond to the eigenvalues of time dependent single particle density matrix expressed in spin-orbitals, taken from TD-FCI.  
Our analysis is motivated by equilibrium, ground-state physics of the underlying half-filled Hubbard model, where natural occupations approach $\lambda_i \rightarrow 0.5$ as the system becomes strongly correlated in the $U\rightarrow \infty$ limit (see Fig.~S3 in supplemental information\cite{supp}).  {More information on the use of natural orbital occupations as a measure of correlation can be found in\cite{Gersdorf_1997,Pulay_1988,Mitxelena2017-ly}. We however emphasize that to our knowledge the use of entropy and natural occupations has not been used as a measure of non-equilibrium induced correlation as we are applying it here.  Crucially, as a result of idempotency the natural occupations of the density matrix for a non-interacting/non-correlated system are either 0 or 1. Furthermore in the supplemental material\cite{supp} we show that this holds even when the system is excited away from equilibrium.  Thus, even in the non-equilibrium regime the natural occupations predict the correct behavior in the limiting case of no correlations making it a reasonable measure of the non-equilbrium induced correlations in the system.}  By examining natural occupations in equilibrium, one can therefore assess how correlated the system is and here we propose this quantity as a measure of correlation in the non-equilibrium setting too.

In Fig.~\ref{fig:entropy} (d), we also look at the von-Neumann entropy for a scan of $E$ values from $1.0J$ to $5.0J$.  The entropy is computed as,
\begin{equation}\label{eq:entropy}
    \mathcal{S}[\rho] = \frac{2}{N}\mathrm{Tr}\left[\rho\log_2(\rho)\right].
\end{equation}
Note that for the ground-state, as $U\to\infty$, the entropy approaches unity.

For $E = 1.0J$, in Fig.~\ref{fig:entropy}(a), we see very little deviation from the equilibrium natural occupations.  This corresponds to near zero entropy of the single-particle density matrix.  Each of our methods matches the benchmark perfectly in equilibrium for the results in Fig.~\ref{fig:diff_E}.  This demonstrates that when the natural occupations remain close to the equilibrium values the four methods capture well  the non-equilibrium TD-FCI results qualitatively and even quantitatively. 

As the excitation strength is increased to $E=2.0J$, we see a more noticeable difference from the equilibrium result.  Though the oscillations of the natural occupations are still relatively small.  The entropy reflects this more clearly where it is only around $\mathcal{S}[\rho] = 0.1$, indicating weak mixing between the natural orbitals.   For the dynamics in Fig.~\ref{fig:diff_E}, the results produced by TD-CC are still in perfect agreement with the TD-FCI result. For the same conditions, the HF-GKBA appear over damped but they still capture elements of the TD-FCI result. 

For $E=5.0J$ we see a strong mixing of the natural occupations, nearing the $\lambda_i[\rho] = 0.5$ limit, corresponding to an entropy of $\mathcal{\mathcal{S}[\rho]}\approx0.9$.  In an equilibrium picture this would correspond to a strongly correlated system, with high $U$. Such a transition was observed for all system sizes and interaction range types. Consequently, we consider this to be a representative case of a system driven by a strong external perturbation from a ``weakly correlated'' ground state to a ``strongly correlated'' excited state.  

Understandably, such a physical regime poses difficulty for methods suited for weak to moderately correlated systems, such as CC or the MBPT expansion. Indeed, TD-CC results fail completely in this regime, as do the KBE results.  In contrast the HF-GKBA offers an improvement in this regime and does a good job at capturing qualitative and quantitative of the TD-FCI results.  In Fig.~S6 of the supplemental information we show the entropy for different $E$ for $N_s=8,12$ and $16$ and see similar behavior for each system size\cite{supp}. We also have noticed that entropy generally increases with system size, with an exception at E=5.0, where we don't observe a maximum entropy for the highest system size we have considered. This presents a competition between field strength and system size to produce the maximally entangled state. We will analyze such behavior in the next publication.

All the 1-RDMs used to generate the time-dependent dipole moments, Von-Neumann entropy, and other results are available on the \texttt{zenodo} repository \cite{zenodo_neq_compare}.  

\section{Discussion and conclusions}\label{sec:discussion}

We now return to discuss the implications of the results we have described in Sec.~\ref{sec:diff_E}-\ref{sec:exciation_analysis}.  Each of the methods we test perform reasonably well in weakly excited systems.  For pulse strengths up to $E=2.0J$ TD-CC gives perfect agreement compared to TD-FCI.
However, as stronger perturbations are applied to drive the system out of equilibrium, TD-CC breaks down. This is because CC wavefunction is expanded around the time-independent, ground-state Hartree-Fock reference which becomes increasingly unsuitable as the system evolves after a strong time-dependent perturbation.

Interestingly, at $E=2.0J$ we observe that TD-HF also improves upon the results for the time-dependent dipole produced by HF-GKBA. Note that this implies that the trajectories of equal-time observables depending on the density matrix (such as the dipole studied here) are qualitatively well captured even by a mean-field technique. {This behavior of the HF-GKBA comes from its ability to redistribute spectral weights into satellite features coming from dynamical self-energy effects. This is why the dynamics begin to flatten, as more of this redistribution occurs the dynamics include more frequencies but all oscillate with reduced amplitude.  This explains why there is an intermediate regime where TD-HF performs better, it is simply a matter of the HF-GKBA redistributing spectral weights incorrectly relative to TD-FCI.  This is also why the benchmark result begins to become closer again to the HF-GKBA result as $E$ is further increased.} This leads to the interesting question of whether in dynamical systems it is always better to include self-energy effects, or whether there are regimes in which the inclusion of the (non-exact) self-energy can have detrimental effects by overdamping the system dynamics as in Fig.~\ref{fig:diff_E} b) or Fig.~\ref{fig:KBE_comparision} b)-d). This is a question that deserves further investigation with additional forms of the self-energy beyond merely the density-density response included in the $GW$ formalism, but also including higher order fluctuations\cite{Joost_2022, vlcek2019stochastic,Mejuto_2022}. {We also would like to highlight other results including recent work investigating artificial damping in NEGF dynamics\cite{Pavlyukh_2024,von_Friesen_2009,von_Friesen_2010,stan_2016}}

{Examining the effect of increasing interaction strength we see that the behavior of the dipole is similar to when we drive the system far from equilibrium.  This suggests a link between increasing correlations due to interaction strength and the possibility that the system is being somehow driven into a strongly correlated regime, even if weakly correlated in equilibrium.  We looked at the effective strength of correlations induced by the non-equilibrium and link the observed performance of the CC and GF methods (within the given approximations) to this. In section \ref{sec:exciation_analysis}, we looked at the different regimes brought about by varying the pulse strength.  The entropy of the density matrix, given in equation \eqref{eq:entropy}, is here used to quantify correlations in the system\cite{Yuan_2022}. In Fig.~\ref{fig:entropy} we showed that for stronger pulses the natural populations become closer to $\lambda=0.5$, which is consistent with those of a strongly interacting equilibrium system as shown in Fig.~S3 of the supplemental information for the model in equation \eqref{eq:MB_ham} with $N_s=6$\cite{supp}.  Further, in in Fig.~\ref{fig:diff_E} (d) the dynamics of the dipole are very slowly varying in time.  This is consistent with the type of dynamics we have seen Fig.~\ref{fig:diff_U} due suppression of the kinetic energy term. We interpret this regime as non-equilibrium induced correlations.}

Interestingly, as we show in Fig.~\ref{fig:diff_U} b),d) and f), when long range interactions are included in the Hamiltonian the result of the HF-GKBA are improved \textit{drastically}.  {This result has also been observed and discussed previously in Ref.~\cite{Reeves_2023} and is attributed to the applied self-energy approximation being a more dominant contribution to the exact self-energy in the long range interacting model and thus capturing the physics better.}

This observation may also help to explain the improvement of the HF-GKBA for $E=5.0J$.  In this highly excited regime, the particles are much more delocalized.  This increases the screening and we propose that this lead to a regime in which the physics accounted for by the self-energy approximation becomes more dominant, leading to reasonably well captured dynamics.  Another reason for the improvement of the HF-GKBA for both $GW$ and SB self-energy approximations compared to TD-HF is that the HF-GKBA allows for the solution to oscillate with more frequencies and reduced amplitudes (due to sattelite solutions) whereas TD-HF is much more limited in how spectral weights can be spread in the spectrum.  {The poor performance of TD-CC is likely arising from the non-hermiticity of the CC equations of motion which leads to more and more unphysical and eventually unstable results for large $E$ or $U$.  }TD-CC can possibly be improved via orbital optimization in the pursuit of renormalizing the interactions, as well as by including additional higher-order excitation terms (beyond the CCSD level).\cite{kvaal_ab_2012,sato_communication_2018,harsha_thermal_cc_2019,pathak_time-dependent_2020}.

From these results, it is clear that the TD-CC performs excellently for relatively weak perturbations and improves upon MBPT, especially in the finer details.  However, in strongly perturbed regimes, TD-CC breaks down and no longer provides results that match TD-FCI, even at the qualitative level.  Notably, for regimes in which the system is strongly perturbed, the HF-GKBA offers results that are very close to those produced by TD-FCI, and are much more numerically stable than those produced by TD-CC.  This points to the possibility that the HF-GKBA and TD-CC have somewhat complimentary regimes of applicability, however definitive statements about this observation will require further exploration of various self-energy formulations.  As is often seen the KBEs show no significant improvement upon the HF-GKBA\cite{Reeves_2023_2}, furthermore in this particular example due to the severe overdamping the HF-GKBA outperforms the KBEs {except in a few cases discussed in Sec.~\eqref{sec:KBE_effects}}. This points to the fact that inclusion of the time-nonlocal components of the self-energy, which dominate the ``quantum memory'' of KBEs, is practically detrimental to the quality of the predicted dynamics, despite its formal correctness. This question warrants further exploration, especially in connection to applications of more involved self-energy formulations, beyond $GW$. 

In this work we have only investigated properties related to the density matrix, namely the dipole and the von~Neumann entropy.  Capturing the dynamics of the density matrix is an important benchmark for any time dependent method, and being able to predict these quantities is valuable for quantitative simulation of carrier dynamics in materials or charge dynamics in quantum chemistry systems.  On the other hand, other measurable quantities of great importance for non-equilibrium quantum physics, such as the the time-dependent spectral function, require knowledge of the time-non-local terms. We are actively investigating the ability of each of these methods to capture the time dependent spectral properties of non-equilibrium systems, as well as looking into efficient ways of improving existing approaches\cite{Reeves_2024}.  

A further extension of this study, also underway, comes in terms of the system sizes studied. Despite the huge reduction in cost afforded in TD-FCI compared to full diagonalization, we are still limited in the system size due to the still large size of the constructed subspace.  One route we are actively pursuing is implementing a time dependent adaptive sampling CI approach(TD-ASCI).  This approach improves upon traditional CI by adding a step that weights wavefunctions making up the subspace, based on estimating how much of a contribution they will make to the subspace.  This can vastly reduce the subspace size thus opening the road to even larger systems.  This presents a route to provide accurate benchmark data in large scale or multi-band periodic systems where such data is extremely difficult to obtain.

In conclusion, the numerical benchmarks presented in this work help to practically assess the performance of variety of time-dependent numerical methods under different strengths of perturbation.  We demonstrate that  TD-CC is an excellent theory for moderate coupling and weak to moderately strong perturbations, but that it becomes unreliable for strong perturbations.  We have further seen the KBEs fail to produce results that improve upon TD-CC.  For weak excitations the HF-GKBA produces worse results than TD-CC but for strong excitations the HF-GKBA gives results that are qualitatively and quantitatively very close to the true result.  Thus, we surmise that a multilayered methodology employing the complementary strengths of TD-CC and HF-GKBA will likely provide accurate results in a wide variety of non-equilibrium regimes.

\section*{Acknowledgements}
This material is based upon work supported by the U.S. Department of Energy, Office of Science, Office of Advanced Scientific Computing Research and Office of Basic Energy Sciences, Scientific Discovery through Advanced Computing (SciDAC) program under Award Number DE-SC0022198.  This research used resources of the National Energy Research Scientific Computing Center, a DOE Office of Science User Facility supported by the Office of Science of the U.S. Department of Energy under Contract No. DE-AC02-05CH11231 using NERSC award BES-ERCAP0029462
\bibliography{Bib}
\end{document}


\preprint{APS/123-QED}

\title{Supplementary information for ``Performance of  wave function and Green’s function methods for non-equilibrium many-body dynamics''}
\author{Cian C. Reeves}
\affiliation{%
Department of Physics, University of California, Santa Barbara, Santa Barbara, CA 93117
}%

\author{Gaurav Harsha}
\affiliation{Department of Chemistry, University of Michigan, Ann Arbor, Michigan 48109, USA}

\author{Avijit Shee}
\affiliation{Department of Chemistry, University of California, Berkeley, USA}

\author{Yuanran Zhu}
\affiliation{Applied Mathematics and Computational Research Division, Lawrence Berkeley National Laboratory,
Berkeley, CA 94720, USA}
\author{Thomas Blommel}
\affiliation{%
Department of Chemistry and Biochemistry, University of California, Santa Barbara, Santa Barbara, CA 93117
}%
\author{Chao Yang}
\affiliation{Applied Mathematics and Computational Research Division, Lawrence Berkeley National Laboratory,
Berkeley, CA 94720, USA}

\author{K Birgitta Whaley }
\affiliation{Department of Chemistry, University of California, Berkeley, USA}
\affiliation{Berkeley Center for Quantum Information and Computation, Berkeley}

\author{Dominika Zgid}
\affiliation{Department of Chemistry, University of Michigan, Ann Arbor, Michigan 48109, USA}
\affiliation{Department of Physics, University of Michigan, Ann Arbor, Michigan 48109, USA}
\author{Vojt\ifmmode \check{e}\else \v{e}ch Vl\ifmmode \check{c}\else \v{c}ek}
\affiliation{%
Department of Chemistry and Biochemistry, University of California, Santa Barbara, Santa Barbara, CA 93117
}%
\affiliation{%
Department of Materials, University of California, Santa Barbara, Santa Barbara, CA 93117
}

\date{\today}
\maketitle

\section{Theory}
\subsection{Time-dependent configuration interaction}

Configuration Interaction (CI) is a wave function-based many-body method, where the exact correlated wave function ($|\Psi \rangle$) is expressed in terms of the ground and excited slater determinants ($|\Phi_I\rangle$)s: $|\Psi \rangle = \sum_I |\Phi_I \rangle$, where I stands for various electronic configurations. When ($|\Psi \rangle$) is plugged into the time-dependent Schr{\"o}dinger equation, we obtain the following equations of motion for the $C_I$s:  

\begin{equation}
    i \frac {\partial C_I}{\partial t} = \sum_J H_{IJ}(t) C_J \label{eq:TDCI}
\end{equation}

The solution of Eqn. \eqref{eq:TDCI} can be written as:
\begin{equation}
C_I (t_n) = \sum _J U_{IJ} (t_n, t_0) C_J(t_0) ; \quad  U_{IJ} (t_n, t_0) = \mathcal{T}\{\exp[-i\int_{t_0}^{t_n} dt H_{IJ}(t)]\} \label{Eq: TDCIsoln}    
\end{equation}

\begin{figure}
    \centering
    \includegraphics[width=.5 \linewidth]{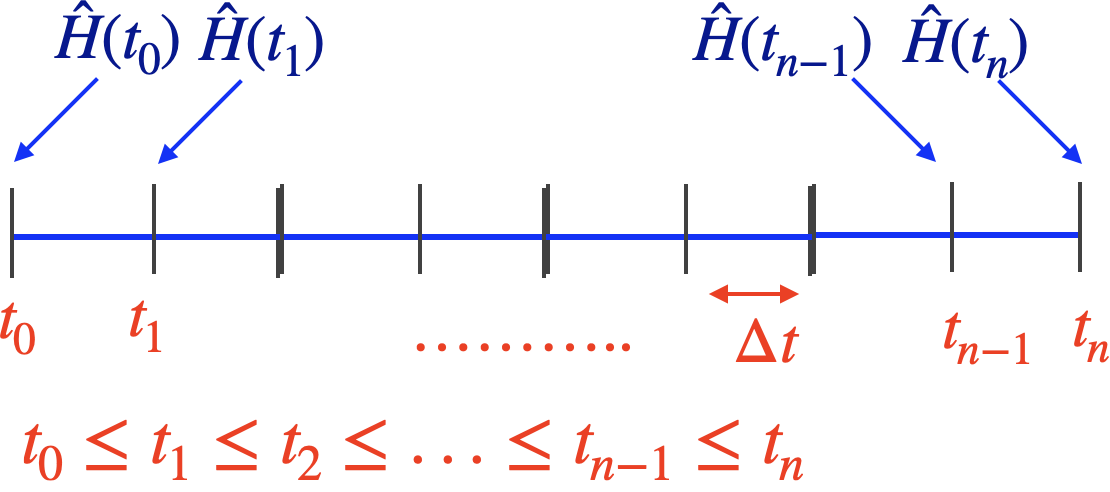}
    \caption{Numerical implementation of time-ordering}
    \label{fig:time_ordering}
\end{figure}

In order to evaluate $U_{IJ} (t_n, t_0)$ numerically we required two components: implementation of the time-ordering and the evaluation of the exponential of a large matrix. In Fig. \ref{fig:time_ordering} we have shown the numerical implementation of the time-ordering. We first break the time interval $[t_0, t_n]$ into many slices and choose a fixed time-ordering: $t_0 \le t_1 \le t_2 \le ...\le t_{n-1} \le t_n$. Also, we assume that the Hamiltonian is piecewise constant within a time-interval: $\hat{H}(t)=\hat{H}(t_p) \forall t_p < t \le t_{p+1}$.

Therefore, Eq. \ref{Eq: TDCIsoln} can be written as:

\begin{equation}
    \mathbf{C}(t_n) = \mathcal{T}\{e^{-i\hat{H}(t_{n-1})\Delta t}....e^{-i\hat{H}(t_1)\Delta t}e^{-i\hat{H}(t_0)\Delta t}\} \mathbf{C}(t_0)
\end{equation}

where we have used the fact that within time-ordering the operators commute. 

To evaluate the exponential of a large matrix (time-independent) we first make a polynomial approximation of the time-evolution operator:

\begin{align}
    C_I (t+\tau) ={}& \sum_J e^{-iH_{IJ} \tau} C_J(t) \\
                 \approx & \sum_J \sum_{k=1}^p \frac {(-it)^k}{k!} H_{IJ}^k C_J(t) \\
                 \approx {}& \sum_{k=1}^p \frac {(-it)^k}{k!} a_I^k(t) \label{Eq:polynomial}
\end{align}

where, $a^k_I (t)$s are 
\begin{equation}
    a^k_I (t) = H_{IJ} a_J^{k-1}(t) ; \quad a^0_J(t) = C_J(t)
\end{equation}

a$^k$ vectors form a subspace $A_k = [a^0 a^1..a^k]$. If we employ the Lanczos procedure to generate the subspace, the Hamiltonian becomes tridiagonal within that subspace. We can do a basis transformation by using the $A_k$ transformation matrix to generate a reduced dimensional (p$\times$p) time-evolution equation \cite{ParkLight_JCP86}. The solution of those equations can be written down as:
 \begin{equation}
     \mathbf{d}(t+\tau) = e^{-iH_p \tau} \mathbf{d}(t) \label{Eq:evolve_Lanczos}
 \end{equation}

Here, $\mathbf{d}(t) = A_k \mathbf{C}(t)$ is a (p $\times$ 1) vector. Finally, $\mathbf{C}$(t+$\tau$)s can be obtained as:

\begin{equation}
    \mathbf{C}(t+\tau) = A_k \mathbf{d}(t+\tau) \label{Eq:krylovtoCI}
\end{equation}

However, the tridiagonal $H_p$ provides a time-local description of the propagator, which becomes unsuitable after a certain time step, $\Delta t^{max}$. This limit can be exactly evaluated \cite{MOHANKUMAR_CPC2006} by analyzing the accuracy of the series expansion up to a threshold $\epsilon$ in Eq. \ref{Eq:polynomial} for a value of p, that is the dimension of the Krylov space. Once, that time limit is reached, we evaluate the CI coefficients at that time-step from Eq. \eqref{Eq:krylovtoCI}, and from there we construct a new subspace time evolution operator, $U_p$. For the time-dependent Hamiltonian, we evaluate the Hamiltonian at (t + $\Delta t^{max}$) and build $U_p$ based on that.  

\paragraph*{Implementaion details} The one- and two-electron integrals for all calculations were prepared using PySCF.~\cite{sun_pyscf_2018, sun_recent_2020} The time-propagation grid in this method is generated by the Lanczos technique, which is typically quite sparse.

\subsection{Time-dependent coupled cluster}
Coupled cluster (CC) theory~\cite{crawford_introduction_2000, bartlett_coupled-cluster_2007} is widely accepted as one of the most accurate wavefunction methods for weakly correlated systems.
This is has led to generalizations of traditional ground-state CC to time-dependent methods.
In this paper, we use the time-dependent CC (TDCC) theory formulated by Arponen,~\cite{arponen_variational_1983} where we first define the time-dependent action integral using the non-Hermitian CC ansatz, such that we have
\begin{equation}
    S = \int dt \mathcal{L}(t) = \int dt \braket{
        \Phi | \left(1 + Z(t)\right) e^{-T(t)}
        \left(
            i \frac{\partial}{\partial t} - H(t)
        \right) e^{T(t)} |
        \Phi
    }.
\end{equation}
Here, $\ket{\Phi}$ is a Hartree-Fock Slater determinant, while the cluster operators $T(t)$ and $Z(t)$ are defined using particle-hole creation and annihilation operators, respectively,
\begin{subequations}
    \begin{align}
        T(t) &= \sum_{ia} \tau_{i}^{a} c^\dagger_a c_i + \frac{1}{4} \sum_{ijab} \tau_{ij}^{ab} c^\dagger_a c^\dagger_b c_j c_i + \cdots,
        \\
        Z(t) &= \sum_{ia} z_i^a c^\dagger_i c_a + \frac{1}{4} \sum_{ijab} z_{ij}^{ab} c^\dagger_i c^\dagger_j c_b c_a + \cdots.
    \end{align}
\end{subequations}
We have adopted chemist's notation for orbital index, i.e., indices $i$, $j$, $k$, $l$, ... denote occupied orbitals, while $a$, $b$, $c$, $d$, ... denote unoccupied (or virtual) orbitals, and finally, $p$, $q$, $r$, $s$, ... are used as general labels.
When the cluster operators are expanded to all orders in particle-hole excitation rank, the CC ansatz provides an exact parametrization of the many-body wave function.
However, for practical implementations, $T$ and $Z$ are truncated to single and double excitation, resulting in the well known CCSD approximation.

The evolution equations for $\tau$ and $z$ amplitudes are found by making the action $S$ stationary with respect to variations in $T$ and $Z$.
In principle, the action also depends on the molecular orbital basis used to construct the mean-field reference $\ket{\Phi}$, such a time-dependent orbital optimization is not considered here.
In other words, $\ket{\Phi}$ is considered to be time-independent and all the time-dependence is incorporated by the CC amplitudes.

Considering the compact notation $T = \sum_\mu \tau_\mu \gamma_\mu^\dagger$ and $Z = \sum_\mu z_\mu \gamma_\mu$, where $\gamma_\mu^\dagger$ and $\gamma_\mu$ represent particle-hole excitation and de-excitation operators, the Lagrangian can be simplified as
\begin{equation}
    \mathcal{L} (t) = i \sum_\mu z_\mu \frac{\partial \tau_\mu}{\partial t} \braket{
        \Phi | \gamma_\mu \gamma_\mu^\dagger | \Phi
    } - \braket{\Phi | (1 + Z) e^{-T} H e^T | \Phi}.
\end{equation}
The associated Euler-Lagrange equations are our desired evolution equations. For the $\tau$-amplitudes, we have
\begin{equation}
    \frac{\partial \mathcal{L}}{\partial z_\mu} = 0
    \Rightarrow
    i \braket{\Phi | \gamma_\mu \gamma_\mu^\dagger | \Phi} \frac{\partial \tau_\mu}{\partial t}
    =
    \braket{
        \Phi | \gamma_\mu e^{-T} H e^T | \Phi
    },
\end{equation}
while for $z$-amplitudes, one obtains
\begin{equation}
    \frac{d}{dt} \left( \frac{\partial \mathcal{L}}{\partial (\partial \tau_\mu / \partial t)} \right)
    = \frac{\partial \mathcal{L}}{\partial \tau_\mu}
    \Rightarrow
    i \braket{\Phi | \gamma_\mu \gamma_\mu^\dagger | \Phi} \frac{\partial z_\mu}{\partial t}
    = - \braket{
        \Phi | (1 + Z) e^{-T} \left[H, \gamma_\mu^\dagger \right] e^{T} | \Phi
    }.
\end{equation}
Explicit expressions for these equations contain a lot of terms, and were derived with the help of {\em drudge} symbolic algebra manipulator.~\cite{zhao_symbolic_2018}
Starting from a known initial condition for the CC amplitudes, these equations can be integrated to obtain the time-dependent CC wave functions.
Unlike traditional ground-state CC, where orbitals and amplitudes can be chosen to be real-valued, in TDCC, the CC amplitudes develop a complex part as we evolve in time.

In CC theory, observable quantities are defined as asymmetric expectation values, i.e., for an operator $A$, we have
\begin{equation}
    \braket{A}_{CC}
    =
    \braket{
        \Phi | (1 + Z) e^{-T} A e^{T} | \Phi
    }.
\end{equation}
Due to the asymmetric nature of this expression, combined with the fact that the $T$ and $Z$ amplitudes are complex numbers, CC estimates for an observable develop a non-physical complex value.
While there may be different ways to address this problem, we adopt the simplest solution by dropping the imaginary part.
It is a general observation that when the CC ansatz is a good approximation, these imaginary components remain small and can be ignored for practical purpose.
On the other hand, when multi-reference character in the time-dependent wave function grows, and CC starts to break down, there is a corresponding increase in the magnitude of both real and imaginary terms in the CC amplitudes.
As a result, CC expectation values return extremely non-physical results; even occupation numbers may become negative or larger than unity.
Therefore, tracking the magnitude of CC amplitudes (and in TDCC, also the size of imaginary part) is a well known practical way to predict the effectiveness of the CC ansatz.
\paragraph*{Implementaion details}The one- and two-electron integrals for all calculations were prepared using PySCF.~\cite{sun_pyscf_2018, sun_recent_2020}
The TD-CC equations for the amplitudes $t_\mu$ and $z_\mu$ are integrated with the Variable-coefficient Ordinary Differential Equation (VODE) solver available in SciPy.~\cite{brown_vode_1989,2020SciPy-NMeth}
The amplitudes are initialized by solving the ground-state coupled cluster equations in the restricted (or symmetry-adapted) formalism.

\subsection{The Kadanoff-Baym equations}

The Kadanoff-Baym equations describe the time evolution of a two-time non-equilibrium GF initially at equilibrium and perturbed by an external field.  In this section, we will introduce the KBEs, as well as some of the theory needed to understand their meaning. 

When a system initially at thermal equilibrium at inverse temperature $\beta$ is perturbed by a time-dependent field, the expectation of an operator can be written generally as\cite{Stan_2009},
\begin{equation}
    \langle \hat{A}(t)\rangle = \frac{\textrm{Tr}[\hat{U}(-i\beta,0)\hat{U}^\dagger(0,t)\hat{A}\hat{U}(t,0)]}{\textrm{Tr}[\hat{U}(-i\beta,0)]}
\end{equation}
\FloatBarrier
\begin{figure}
    \centering
    \includegraphics[width=\linewidth]{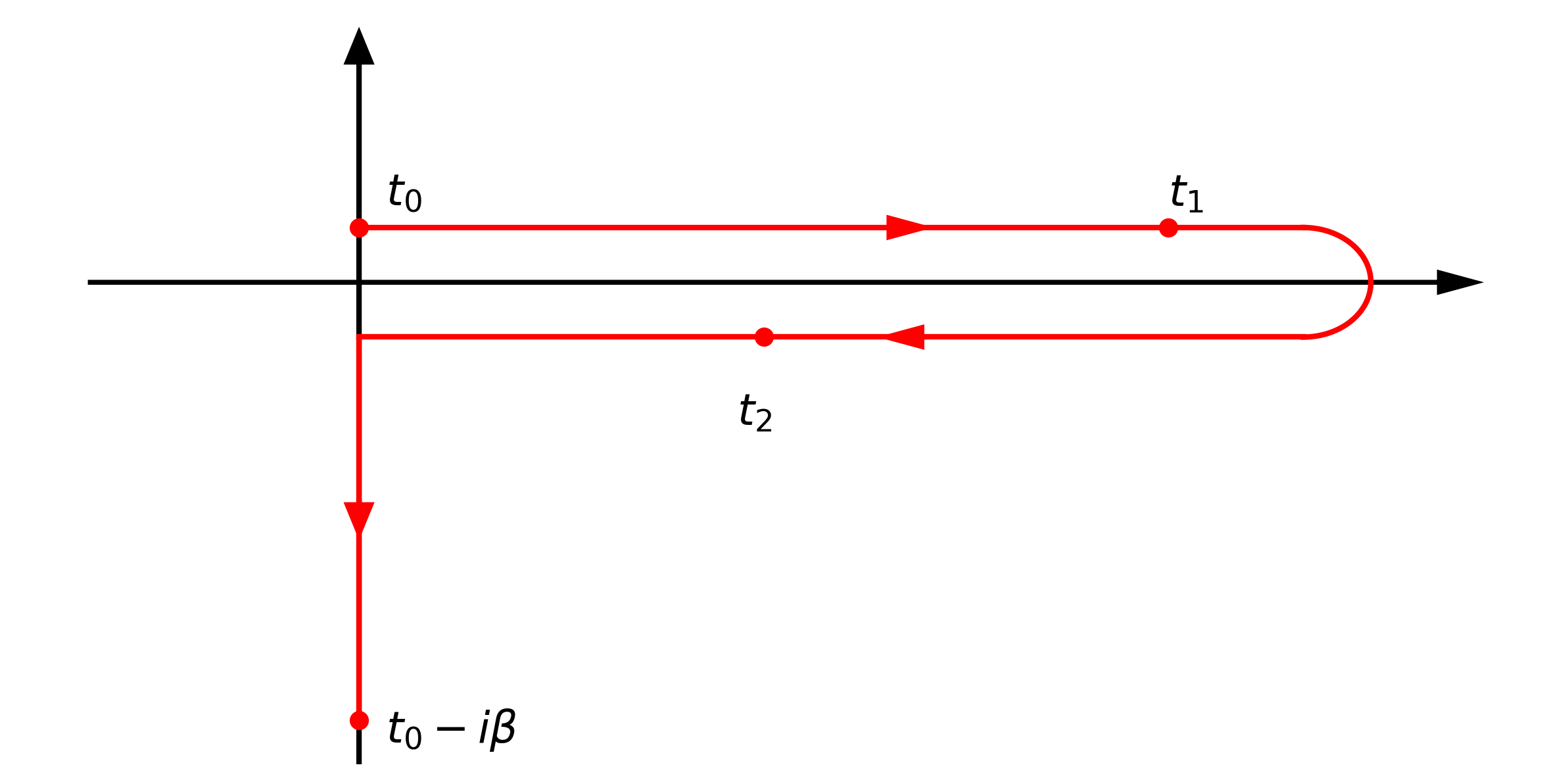}
    \caption{The Keldysh contour.  The contour time ordering operator, $\mathcal{T}_C$, places operators from right to left such that their time arguments follow the order of the arrows shown in the contour.  The vertical portion of the track relates to the initial preparation while the horizontal portion is related to the non-equilibrium evolution of the operator.}
    \label{fig:keldysh_contour}
\end{figure}
\noindent Here, without loss of generality, we assume the time-dependent field is switched on at $t=t_0=0$.  In the above we also have,
\begin{equation}
\begin{split}
    U(t_2,t_1)&=
    \begin{cases}
    \mathcal{T}\{\exp[-i\int_{t_1}^{t_2} dt\mathcal{H}(t)]\} \quad t_2 > t_1\\
    \Bar{\mathcal{T}}\{\exp[-i\int_{t_1}^{t_2} dt\mathcal{H}(t)]\} \quad t_2 < t_1,
    \end{cases}\\
\end{split}
\end{equation}
where $\mathcal{T} (\bar{\mathcal{T}})$ is the (anti-)time-ordering operator.  Finally,
\begin{equation}
        U(-i\beta,0) = \exp(-\beta \mathcal{H}(0)),
\end{equation}
is the thermal statistical averaging operator for a given inverse temperature $\beta$. In the following sections, the finite temperature formalism at a sufficiently low temperature is invoked in preparing the correlated ground state for the KBEs. In contrast, the strictly 0K formalism typically employs adiabatic connection to prepare the correlated quantum state. Within the Keldysh formalism, this product can equivalently be written as an ordered product on the contour shown in Fig. \ref{fig:keldysh_contour}.  The operator $\mathcal{T}_C$ in equation \eqref{eq:contour_G} denotes the contour ordering and places operators from right to left with time arguments in the order that corresponds to the direction of the arrows appearing in Fig. \ref{fig:keldysh_contour}. While the details of the Keldysh formalism are not crucial in this work, we do introduce some common definitions in order to present the Kadanoff-Baym equations for the NEGF. For more information on the Keldysh formalism, we direct the reader to \cite{stefanucci2013nonequilibrium,Kadanoff_1962,bonitz2015quantum,Stan_2009}. 

Firstly, in the Keldysh formalism the single particle GF can be written as,
\begin{equation}\label{eq:contour_G}
    G(t_1,t_2) = -i \langle \mathcal{T}_C[c(t_1)c^\dagger(t_2)]\rangle.
\end{equation}
At inverse temperature $\beta$ and with $t_0$ denoting the time at which the system leaves equilibrium, $t_1$ and $t_2$ can lie anywhere on the contour shown in Fig. \ref{fig:keldysh_contour}.  When both time arguments lie on the real axis the lesser and greater GFs are defined as follows,
\noindent
\begin{equation}
    \begin{split}
        G^{>}(t_1,t_2) &= -i\langle c(t_1) c^\dagger(t_2)\rangle,\\
        G^{<}(t_1,t_2) &= i\langle  c^\dagger(t_2)c(t_1)\rangle.
    \end{split}
\end{equation}
The time ordering operator means the full GF in equation \eqref{eq:contour_G} will be given by $G^{>}(t_1,t_2)$ when $t_1$ is later than $t_2$ and $G^{<}(t_1,t_2)$ when $t_2$ is later than $t_1$ and both are on the real axis.  When one of the time arguments lies on the imaginary axis we have,
\begin{equation}
\begin{split}
        G^{\lceil}(-i\tau,t) &= G^{>}(t_0-i\tau,t),\\
        G^{\rceil}(t,-i\tau) &= G^{<}(t,t_0-i\tau).\\
\end{split}
\end{equation}
Finally, when both arguments lie on the imaginary axis we are left with the time translation invariant Matsubara GF,
\begin{equation}
    iG^\mathrm{M}(\tau_1-\tau_2) = G(t_0 - i\tau_1,t_0 - i\tau_2),
\end{equation}
which represents the equilibrium GF at a given temperature. The same definitions also exist for the self energy operator.
Using these notations the KBEs for time propagation of the NEGF can be written explicitly as,
\begin{equation}\label{eq:KBE}
           \begin{split}
                [-\partial_\tau - h] G^\mathrm{M}(\tau) &= \delta(\tau) + \int_0^\beta d\Bar{\tau}\Sigma^\mathrm{M}(\tau-\Bar{\tau})G^\mathrm{M}(\bar{\tau}),\\
                i\partial_{t_1} G^{\rceil}(t_1,-i\tau) &= h^{\textrm{HF}}(t_1)G^{\rceil}(t_1,-i\tau) + I^{\rceil}(t_1,-i\tau),\\
                -i\partial_{t_2} G^{\lceil}(-i\tau,t_2) &= G^{\lceil}(-i\tau,t_2)h^{\textrm{HF}}(t_2) + I^{\lceil}(-i\tau,t_1),\\
                i\partial_{t_1} G^{\lessgtr}(t_1,t_2) &= h^{\textrm{HF}}(t)G^{\lessgtr}(t_1,t_2) + I_1^{\lessgtr}(t_1,t_2),\\
                -i\partial_{t_2} G^{\lessgtr}(t_1,t_2) &= G^{\lessgtr}(t_1,t_2)h^{\textrm{HF}}(t_2) + I_2^{\lessgtr}(t_1,t_2),\\
            \end{split}
\end{equation}
with the so-called collision integrals being given by

\begin{equation}\label{eq:coll_int}
    \begin{split}
        I_{1}^{\lessgtr}(t_1,t_2) &= \int_{0}^{t_1}\mathrm{d}\bar{t} \Sigma^\mathrm{R}(t_1,\Bar{t})G^{\lessgtr}(\Bar{t},t_2) +\int_{0}^{t_2} \mathrm{d}\bar{t} \Sigma^{\lessgtr}(t_1,\Bar{t})G^\mathrm{A}(\Bar{t},t_2)- i\int_0^\beta \mathrm{d}\bar{\tau} \Sigma^\rceil(t_1,-i\bar{\tau})G^{\lceil}(-i\bar{\tau},t_2)\\
        I_{2}^{\lessgtr}(t_1,t_2) &= \int_{0}^{t_1} \mathrm{d}\bar{t} G^\mathrm{R}(t_1,\Bar{t})\Sigma^{\lessgtr}(\Bar{t},t_2) + \int_{0}^{t_2} \mathrm{d}\bar{t} G^{\lessgtr}(t_1,\Bar{t})\Sigma^\mathrm{A}(\Bar{t},t_2) - i\int_0^{\beta}\mathrm{d}\bar{\tau}G^{\rceil}(t_1,-i\bar{\tau})\Sigma^{\lceil}(-i\bar{\tau},t_2)\\
        I^{\rceil}(t_1,-i\tau) &= \int_{0}^{t_1} d\bar{t} \Sigma^\mathrm{R}(t_1,\bar{t})G^{\rceil}(\bar{t},-i\tau) + \int_0^\beta d\bar{\tau} \Sigma^{\rceil}(t_1,-i\bar{\tau})G^M(\bar{\tau} - \tau),\\
        I^{\lceil}(-i\tau,t_1) &= \int_0^{t_1}d\bar{t} G^{\lceil}(-i\tau,\bar{t})\Sigma^\mathrm{A}(\bar{t},t) +\int_0^{\beta}d\bar{\tau}G^\mathrm{M}(\tau - \bar{\tau}) \Sigma^{\lceil}(-i\bar{\tau},t_1).
    \end{split}
\end{equation}
Here the retarded/advanced Green's function $G^{\mathrm{R/A}}$ and self energy $\Sigma^{\mathrm{R/A}}$ are functions of $G^{\lessgtr}$. In the above equations, the self energy $\Sigma(t_1,t_2)$ includes only correlation terms, while the Hartree-Fock contribution is included in $h^{\mathrm{HF}}(t)$. The first equation in \eqref{eq:KBE} describes the role of the initial correlations in the propagation of the NEGF. The remaining equations describe the propagation of the two-time particle and hole propagator after leaving equilibrium. 

The two-time nature of the KBEs combined with the various collision integrals means the cost of solving these equations scales cubically in the number of time steps.  Several approaches have been employed to circumvent the difficulty of performing long NEGF time evolutions.  This includes extrapolation of trajectories from a short snapshot of the initial dynamics as well reducing cost through stochastic compression of matrix contractions\cite{Reeves_2023,Yin_2021,mejía_2023} . Another approach is through direct approximation of the full KBEs.  The most popular of these approximation schemes is known as the Hartree-Fock generalized Kadanoff-Baym ansatz (HF-GKBA).  In the following section, we will introduce the HF-GKBA.

\paragraph*{Implementation details}The KBEs are solved using the NESSi simulation library\cite{Schuler_2020} using a time step of $dt = .025$ and an inverse temperature $\beta = 20$. This value of $\beta$ was chosen by converging the dynamics with respect to $\beta$, so that this effectively is a zero temperature simulation.
\subsection{The Generalized Kadanoff-Baym Ansatz}
Now we will discuss a commonly used approximation to the Kadanoff-Baym equations, known as the Hartree-Fock generalized Kadanoff-Baym ansatz.  Unlike the KBEs, the HF-GKBA does not prepare initial correlations through the contour integration but rather through other means such as adiabatic switching\cite{Touvinen_2019,Karlsson_2018}. For this reason only the KBEs with real time arguments are considered in the derivation of the GKBA.  The equation of motion for the time-diagonal Green's function can be derived by combining the final two equations in equation \eqref{eq:KBE},
\begin{equation}\label{eq:time_diagonal}
           \begin{split}
                i\partial_t G^{<}(t,t) &= [h^{\textrm{HF}}(t),G^{<}(t,t)] + I_1^<(t,t) - I_2^{<}(t,t)
            \end{split}
        \end{equation}
with 
\begin{equation}\label{eq:coll_int_GKBA}
    \begin{split}
        I_{1}^{\lessgtr}(t_1,t_2) &= \int_{0}^{t_1} \mathrm{d}\bar{t} \Sigma^\mathrm{R}(t_1,\Bar{t})G^{\lessgtr}(\Bar{t},t_2) +\int_{0}^{t_2} \mathrm{d}\bar{t} \Sigma^{\lessgtr}(t_1,\Bar{t})G^\mathrm{A}(\Bar{t},t_2)\\
        I_{2}^{\lessgtr}(t_1,t_2) &= \int_{0}^{t_1} \mathrm{d}\bar{t} G^\mathrm{R}(t_1,\Bar{t})\Sigma^{\lessgtr}(\Bar{t},t_2) + \int_{0}^{t_2} \mathrm{d}\bar{t} G^{\lessgtr}(t_1,\Bar{t})\Sigma^\mathrm{A}(\Bar{t},t_2).
    \end{split}
\end{equation}
This form of the collision integrals assumes the state at $t=0$ has already been prepared in the correlated ground state.  We note that now the time arguments lie strictly on the real-time axis. 

The HF-GKBA is derived directly from the KBE and can be summarized in the following equations\cite{Hermanns_2012},
\begin{equation}\label{eq:HF-GKBA}
   \begin{split}
        G^{\lessgtr}(t_1,t_2) &= iG^\mathrm{R}(t_1,t_2)G^{\lessgtr}(t_2,t_2) - iG^{\lessgtr}(t_1,t_1)G^\mathrm{A}(t_1,t_2),\\
        G^{\mathrm{R,A}}(t_1,t_2)&=\pm \Theta[\pm(t_1 - t_2)]T\{\mathrm{e}^{-i\int_{t_2}^{t_1} h^{\textrm{HF}}(t) dt}\}.
   \end{split}
\end{equation}
In other words, at each time step only equation \eqref{eq:time_diagonal} is explicitly evaluated. Equation \eqref{eq:HF-GKBA} is then used to reconstruct the time off-diagonal components.

Apart from those approximations made to the self energy, which HF-GKBA and KBE share,  two additional approximations are made in the derivation of HF-GKBA.  The first involves neglecting certain integrals that account for time non-local memory effects.  These terms appear in the expression for reconstructing $G^{\lessgtr}(t,t')$.  Once dropped, one is left with the first expression in equation \eqref{eq:HF-GKBA}.  With no further approximation, this ansatz for the time off-diagonal components of $G^{\lessgtr}(t,t')$ is referred to as the generalized Kadanoff-Baym ansatz(GKBA)\cite{Lipavsky_1986}.   The HF-GKBA involves a further approximation where the full $G^{\mathrm{R/A}}(t,t')$ are replaced by the retarded and advanced Hartree-Fock propagator.   Notably, the HF-GKBA leaves important quantities such as energy and particle number conserved as well as retaining causal time evolution.

Recently a linear time scaling$[\sim O(N_t)]$ implementation of the HF-GKBA has been achieved, opening the door for long-time evolution of NEGFs\cite{Joost_2020}. The method removes the explicit appearance of integrals in equation \eqref{eq:coll_int_GKBA} from the differential equation for $G^<(t)$ by expressing them in terms of the correlated part of the equal time two-particle GF $\mathcal{G}(t)$. Within this formulation $\mathcal{G}(t)$ is propagated simultaneously with $G^<(t)$ using an expression analogous to equation \eqref{eq:time_diagonal}.  

The exact equation of motion for $\mathcal{G}(t)$ depends on the self energy approximation used. In this work we use the $GW$ self-energy, in whihc case

For the $GW$ self energy, the equations of motion for $G^<(t)$ and $\mathcal{G}(t)$ in the orbital basis are given below\cite{Schlunzen_2020}.
\begin{equation}\label{eq:G1-G2}
\begin{split}
       i \partial_t G^<_{ij}(t) &= [h^{\textrm{HF}}(t), G^<(t)]_{ij} + [I+I^\dagger]_{ij}(t)\\   i\partial_t \mathcal{G}_{ijkl}(t) &= [h^{(2),\textrm{HF}}(t),\mathcal{G}(t)]_{ijkl} +\Psi_{ijkl}(t) + \Pi_{ijkl} - \Pi_{lkji}^*.
\end{split} 
\end{equation}
Above, the following definitions are made,
\begin{align}
        h_{ij}^{\textrm{HF}}(t) &= h^{(0)}_{ij}(t) - i\sum_{kl} [2w_{ikjl}(t) - w_{iklj}(t)]G_{kl}^<(t)\nonumber,\\
        I_{ij}(t)&=-i\sum_{klp} w_{iklp}(t)\mathcal{G}_{lpjk}(t),\nonumber\\
        h^{(2),\textrm{HF}}_{ijkl}(t) &= \delta_{jl}h^{\textrm{HF}}_{ik}(t) + \delta_{ik}h^{\textrm{HF}}_{jl}(t)\nonumber,\\  
    \Psi_{ijkl} &=\sum_{pqrs}[w_{pqrs}(t) - w_{pqsr}(t)]\times\nonumber \bigg[G^>_{ip}(t) G^<_{rk}(t) G^>_{jq}(t)G^<_{sl}(t)- G^<_{ip}(t)G^>_{rk}(t)G^<_{jq}(t)G^>_{sl}(t)\bigg],\\
    \Pi_{ijkl} &= \sum_{pqrs}w_{rqsp}\left[G_{jr}^>G_{sl}^< - G_{jr}^<G_{sl}^>\right]\mathcal{G}_{ipkq}
\end{align}
Here $h^{(0)}(t)$ is the single particle Hamiltonian and $w_{ijkl}(t)$ is the two-body interaction matrix.   The time dependence given to $w_{ijkl}$ is to allow for adiabatic switching for preparation of the initial state.   The tensor $\Psi_{ijkl}$ accounts for pair correlations built up due to two-particle scattering events and $\Pi_{ijkl}$ accounts for polarization effects\cite{Joost_2020}.

\paragraph*{Implementation details:} The TD-HF and HF-GKBA equations of motion are solved with the 4th order Runge-Kutta algorithm using a time step of $dt = 0.02$. The TD-HF calculation is the special case of $\mathcal{G}(t)= 0$ in equation \eqref{eq:G1-G2}.

\section{Additional Results}
\subsection{Natural populations as a measure of correlation}
\begin{figure}
    \centering
    \includegraphics[width=\textwidth]{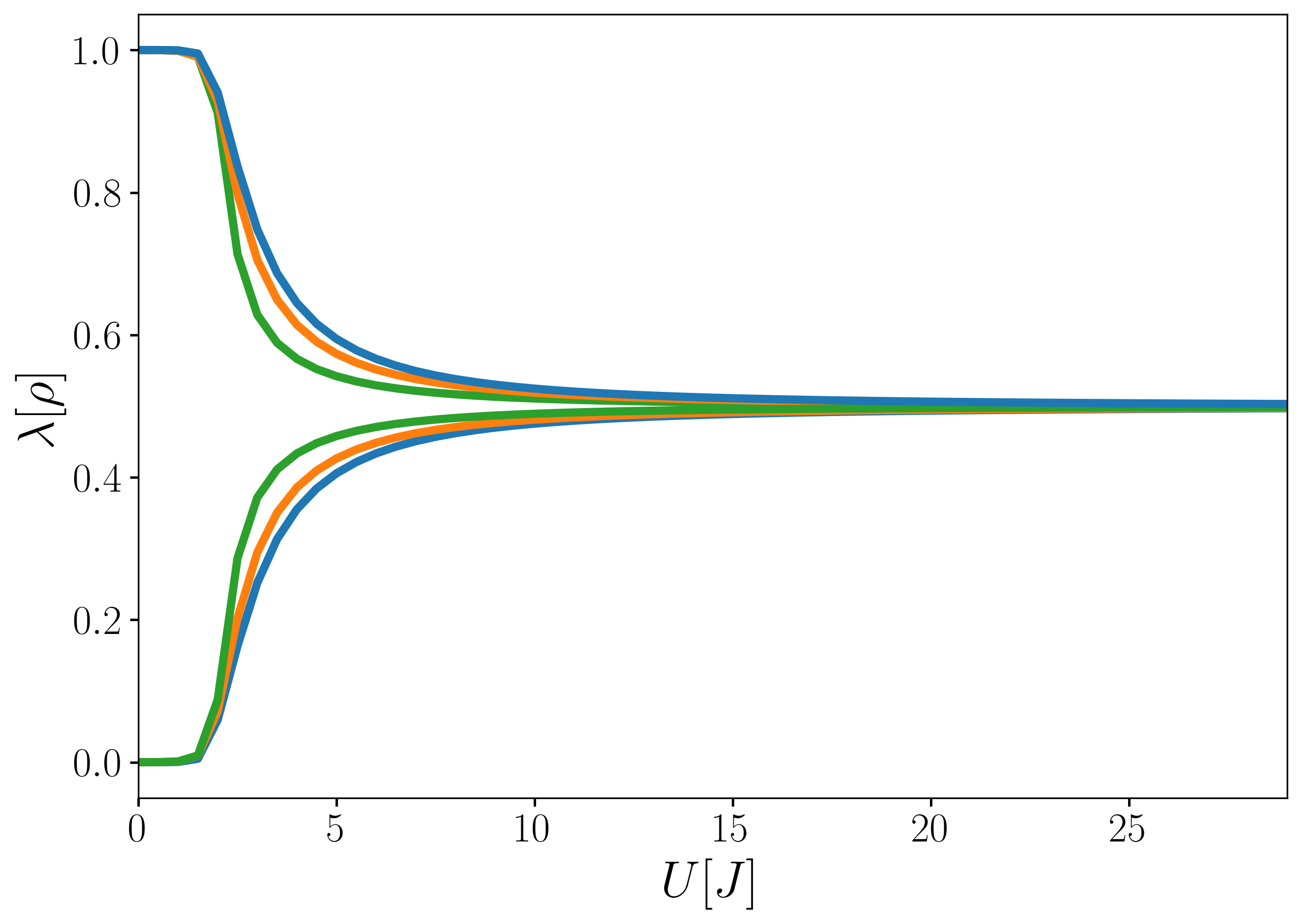}
    \caption{Example showing the natural occupations as a function of $U$ for the model given in equation (6) in the main text with $N_s=6$.  This helps demonstrates the link between the natural occupations and strong interactions or correlations.  In the limit $U\rightarrow\infty$ the natural occupations approach $0.5$.  In the main text we look at the evolution of the natural occupations as a measure of the effect of driving on the correlations in the system, there we see that for large field strengths the natural occupations also approach $0.5$}
    \label{fig:nat_occs}
\end{figure}
\FloatBarrier

\subsection{Effects of long range interactions}
\begin{figure}
    \centering
\includegraphics[width=\textwidth]{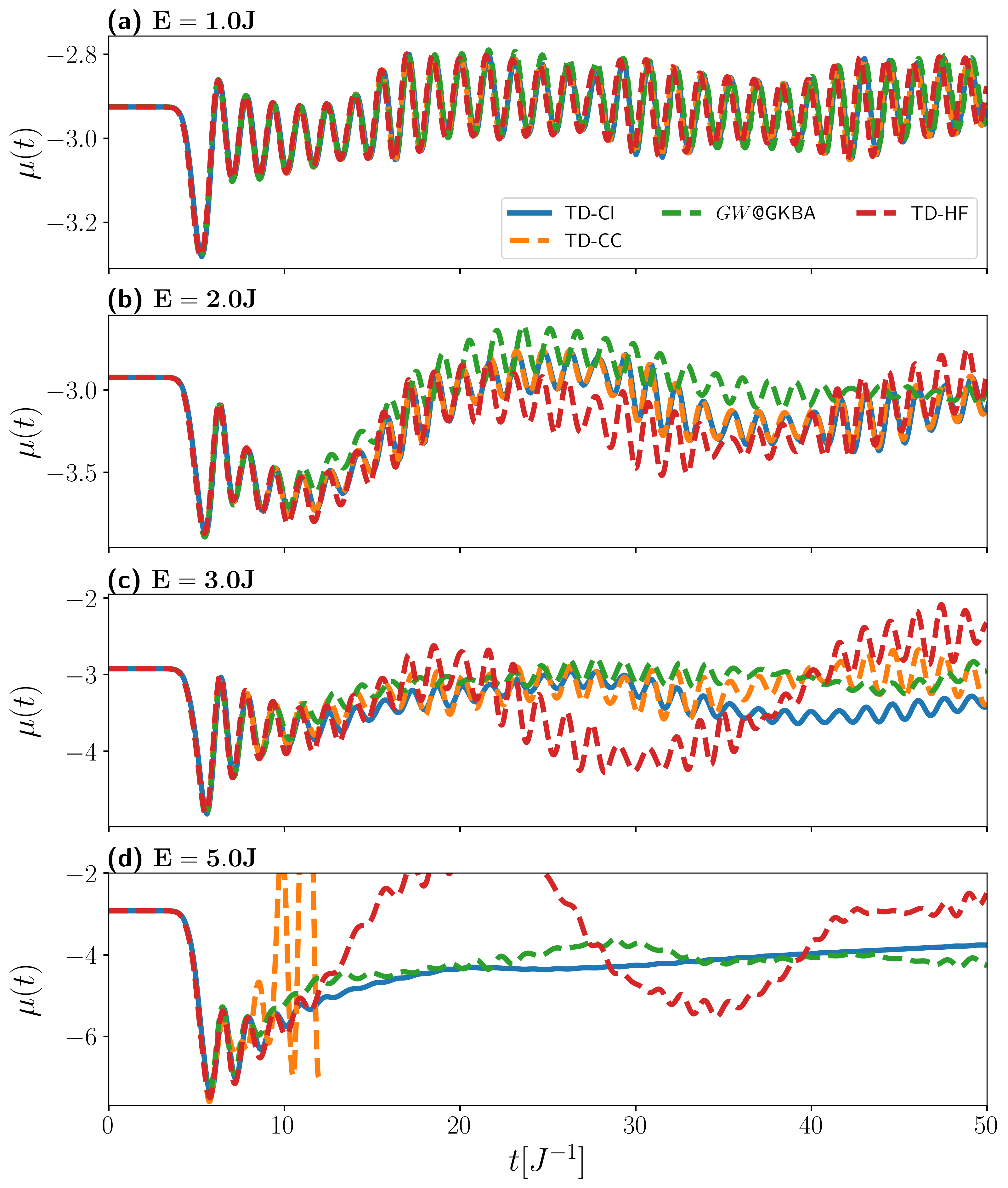}
    \caption{Dynamics for the system described in equation (6) and (7) of main text for different values of the electric field strength $E$ and with $U = 1.0J$ and $\gamma = 0.5$. a) $E=1.0J$, b) = $2.0J$, c) $4.0J$, d) $5.0J$}
    \label{fig:long_range}
\end{figure}

In Fig. \ref{fig:long_range} we show results for a long range interacting model with different $E$ values with $U=1.0$ and $\gamma=0.5$.
\subsection{Additional data for varying $U$ and $E$ values}
\begin{figure}
    \centering
    \includegraphics[width=\textwidth]{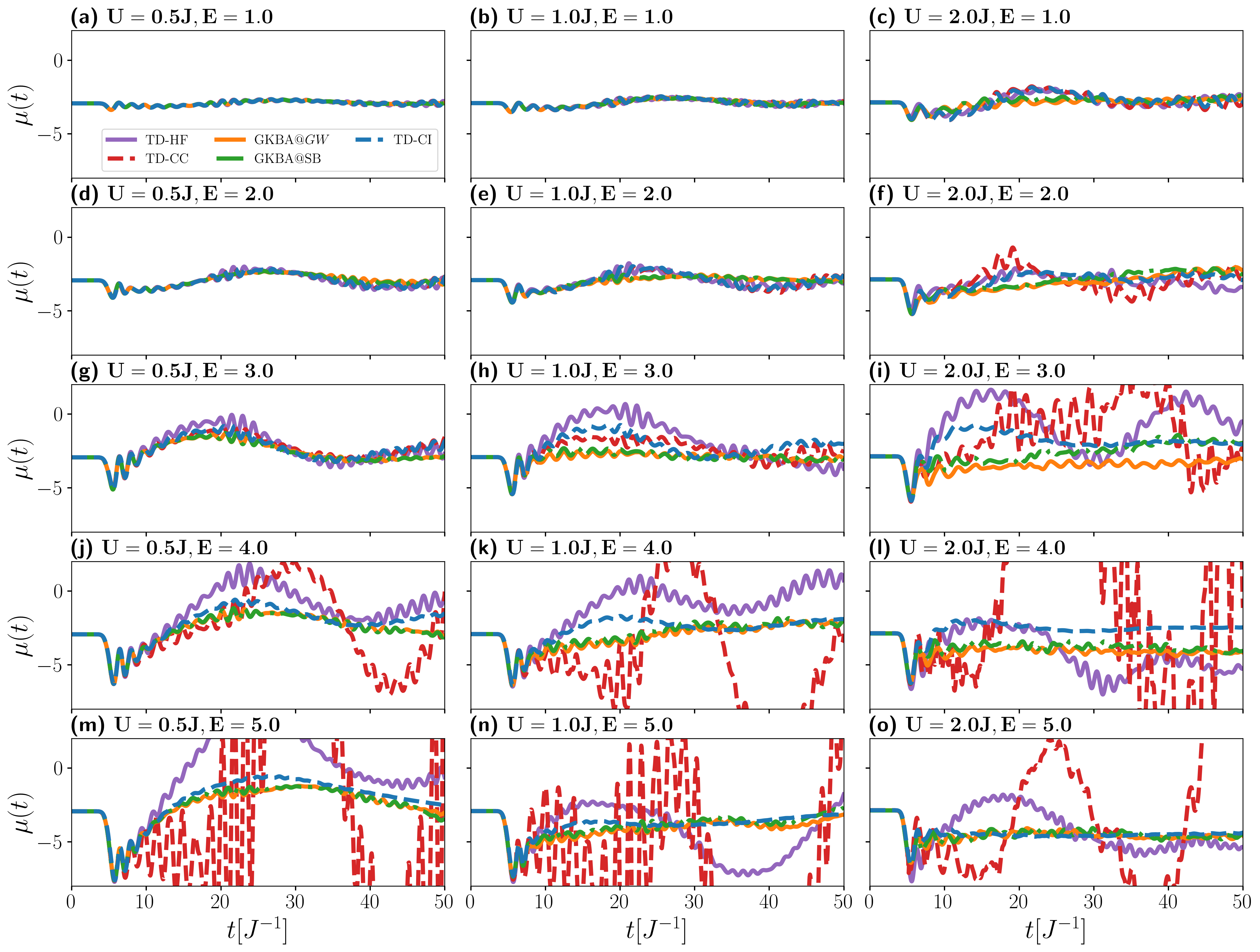}
    \caption{Dynamics of the electronic dipole moment for the generalized Hubbard model ((6) with $\gamma =0$) for all combinations of $U=0.5J,1.0J,2.0J$ and $E=1.0J,2.0J,3.0J,4.0J$ and $5.0J$}
    \label{fig:fullset}
\end{figure}

In Fig. \ref{fig:fullset} we show results for all combinations of $U$ and $E$ investigated for the onsite interacting model.  
\subsection{Entropy for different system sizes}
\begin{figure}
    \centering
\includegraphics[width=\textwidth]{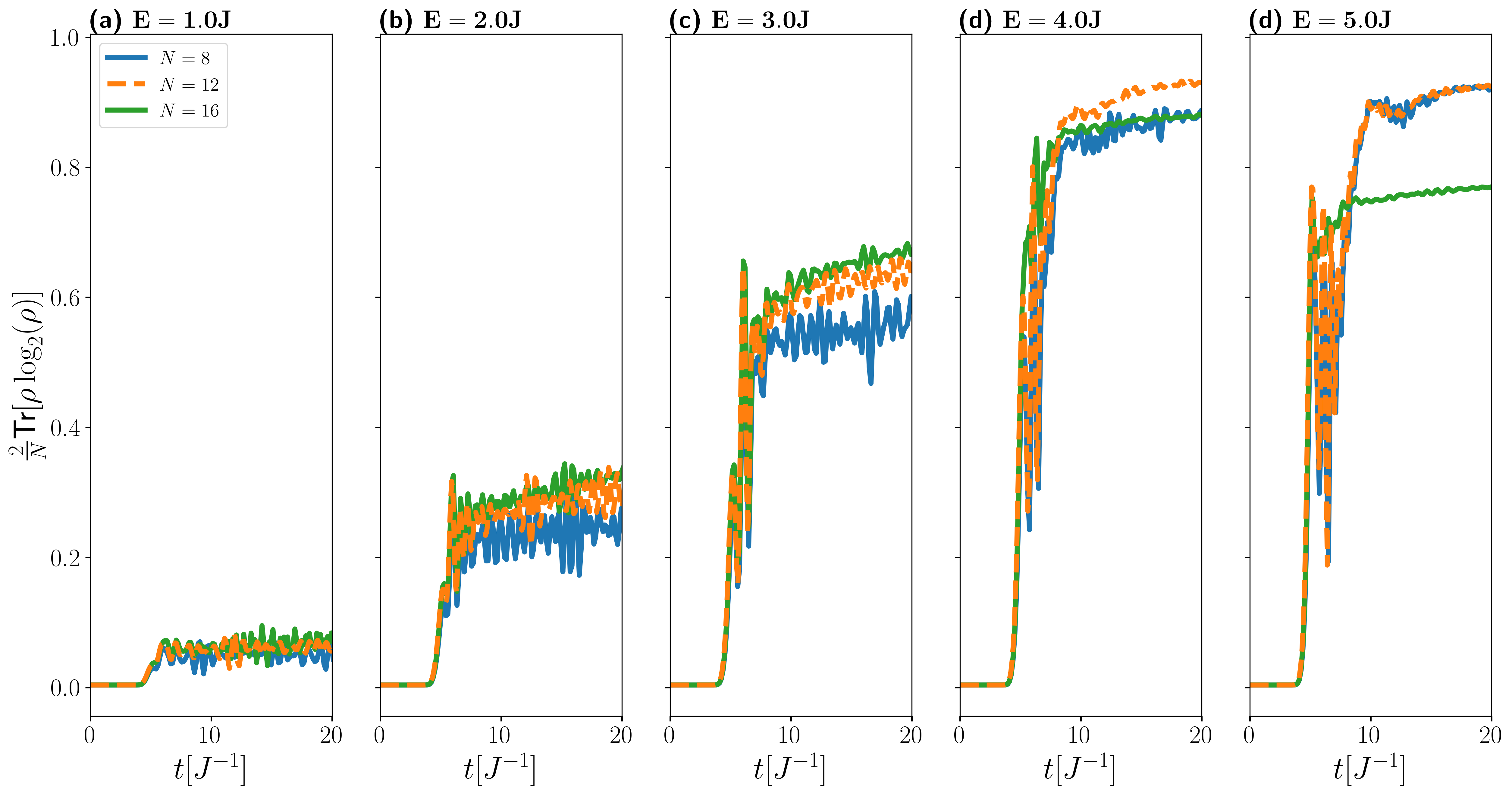}
    \caption{Von-neumann entropy evaluated with TD-CI for system sizes $N_s=8,12$ and $16$ with onsite interactions.  The system is perturbed from it's ground state by the pulse in equation (7) in the main text.  Each panel corresponds to a different value of the external field strength $E$.  a) $E=1.0J$, b) $E=2.0J$, c) $E=3.0J$, d) $E=4.0J$, e) $E=5.0J$  }
\label{fig:size_dependence}
\end{figure}

In this section we provide entropy calculations for different system sizes using TD-FCI. The results are shown in Fig. \ref{fig:size_dependence}

\subsection{Imaginary parts in CC occupation numbers}
\begin{figure}
    \centering
    \includegraphics[width=\textwidth]{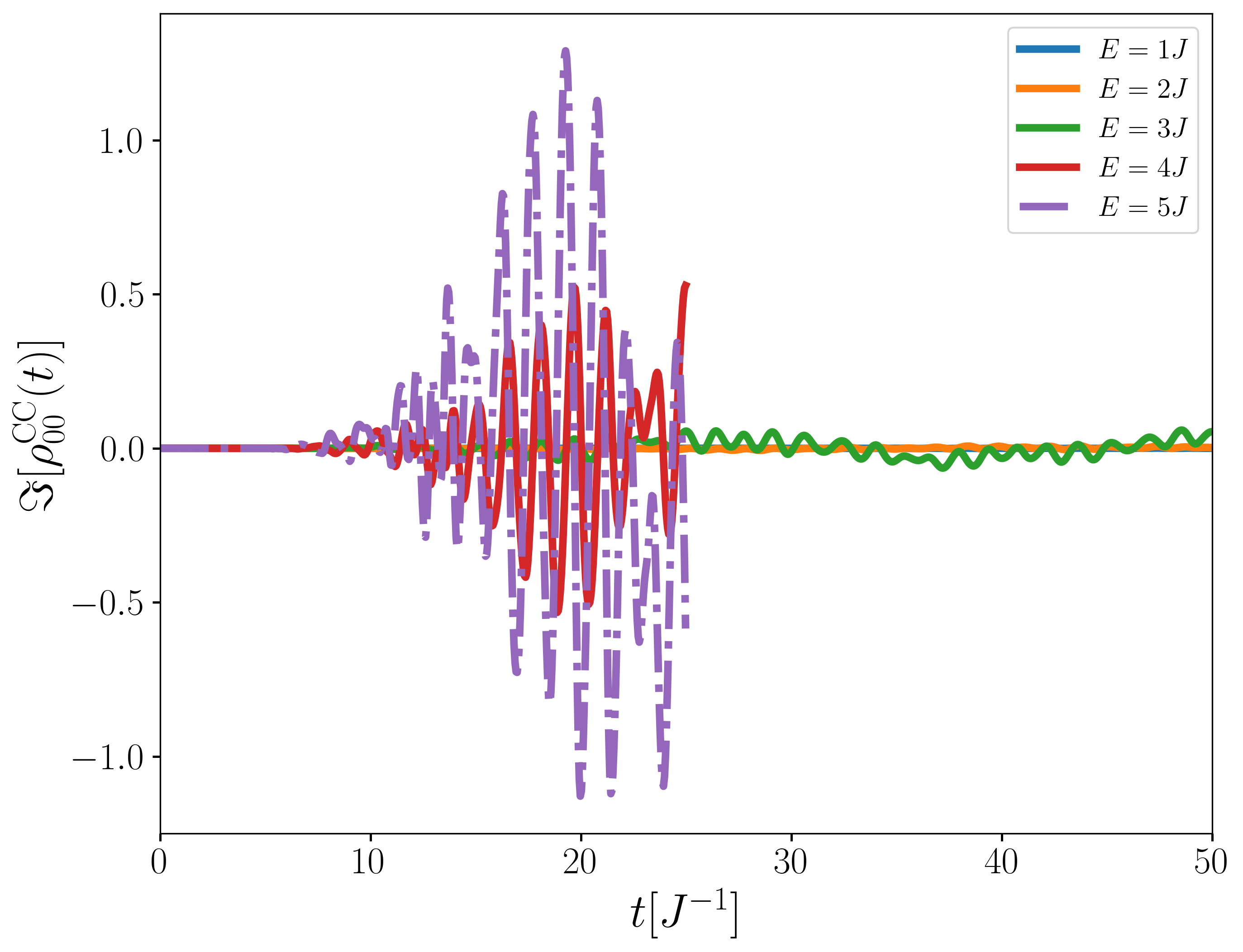}
    \caption{Imaginary part in occupation number for the first site in the model lattice in equation (7) in the main text, as predicted by time-dependent CCSD for $N_s=12$. Results for different perturbation strength $E$ show that for large perturbation, TD-CC occupations develop imaginary parts, that increase as we evolve in time. On the other hand, for weak interactions, the imaginary terms are well behaved and can be disregarded.}
    \label{fig:imag_occ_cc}
\end{figure}

In the main text, we pointed out that TD-CC, due to its non-Hermitian nature, develops imaginary or unphysical parts in physical observables.
In Fig.~\ref{fig:imag_occ_cc}, as a prototypical example, we plot the imaginary parts for the TD=CC expectation value for electron density on the first site in Hubbard lattice.
As the strength of the time-dependent perturbation increases, TD-CC results start accumulating increasingly larger imaginary parts in the occupation number.
In fact, for $E=4J$ and $E=5J$, with such large unphysical components in electron density, the theory is no longer well behaved, and can be deemed a failure.
This property, however, makes TD-CC unique as it provides a diagnostic tool to assess its own effectiveness.
\FloatBarrier
\section{Proof of conservation of idempotency in non-interacting system}
If we assume our system to be non-interacting we have the following idempotency property for the initial density matrix
\begin{equation*}
    \rho^2(0) - \rho(0) = 0
\end{equation*}
Using this we will show that $\rho^2(t) - \rho(t) = 0$ and thus that $\rho$ remains idempotent under unitary time-evolution.  For a non-interacting system the time-evolution of the single particle density matrix is given by $\rho(t) = U(t)\rho(0)U^\dagger(t)$ where $U(t)$ is the single particle time-evolution operator.  From this we have,
\begin{equation*} 
    \begin{split}
        \rho^2(t) - \rho(t)&=\\
U(t)\rho(0)\underbrace{U^\dagger(t)U(t)}_{\mathbf{I}}\rho(0)U^\dagger(t) - U(t)\rho(0)U^\dagger(t)&=\\
        U(t)\rho(0)\rho(0)U^\dagger(t) - U(t)\rho(0)U^\dagger(t)&=\\
        U(t)\rho(0)^2U^\dagger(t) - U(t)\rho(0)U(t)&=\\
        U(t)\rho(0)U^\dagger(t) - U(t)\rho(0)U(t)&=0.
    \end{split}
\end{equation*}
Here we have used the unitarity of the time-evolution operator to go from line 2 to line 3 as well as the idempotency of the initial density matrix in going from line 4 to line 5. Thus, even if the system is highly excited it will remain idempotent.
\bibliography{Bib2}